\documentclass[10pt,twocolumn,letterpaper]{article}

\usepackage[pagenumbers]{cvpr} 

\definecolor{cvprblue}{rgb}{0.21,0.49,0.74}
\usepackage[pagebackref,breaklinks,colorlinks,allcolors=cvprblue]{hyperref}

\usepackage{xcolor}
\usepackage{booktabs}
\usepackage{colortbl}
\usepackage{multirow, multicol}
\usepackage{algorithm, algorithmic}
\usepackage{amssymb}
\usepackage{graphicx}
\usepackage{makecell}
\usepackage{xspace}

\definecolor{custompink}{RGB}{255, 105, 180}

\newcommand{\ours}{RG}
\newcommand{\secs}{\,s\xspace}
\newcommand{\oursrow}{\rowcolor{gray!10}}
\newcommand{\spanval}[1]{\multicolumn{2}{c}{[\hfill #1\hfill]}}
\definecolor{skyblue}{RGB}{60,140,200}

\newcommand{\yold}{y_{\text{old}}}
\newcommand{\ynew}{y_{\text{new}}}

\def\confName{CVPR}
\def\confYear{2027}

\title{Residual Guidance for Real-Time Interactive Video Generation}

\author{NaHyeon Park$^{1}$ \quad Kyeong-Hwan Moon$^{2}$ \quad Youjin Jeon$^{1}$ \quad Hyunjung Shim$^{1}$\\
$^{1}$KAIST \qquad $^{2}$POSTECH\\
{\hypersetup{urlcolor=skyblue}Project page: \url{https://residual-guidance.github.io}}
}

\makeatletter
\let\orig@maketitle\@maketitle
\def\@maketitle{%
  \orig@maketitle
  \vspace{-22pt}
  \begin{center}
    \includegraphics[width=\textwidth]{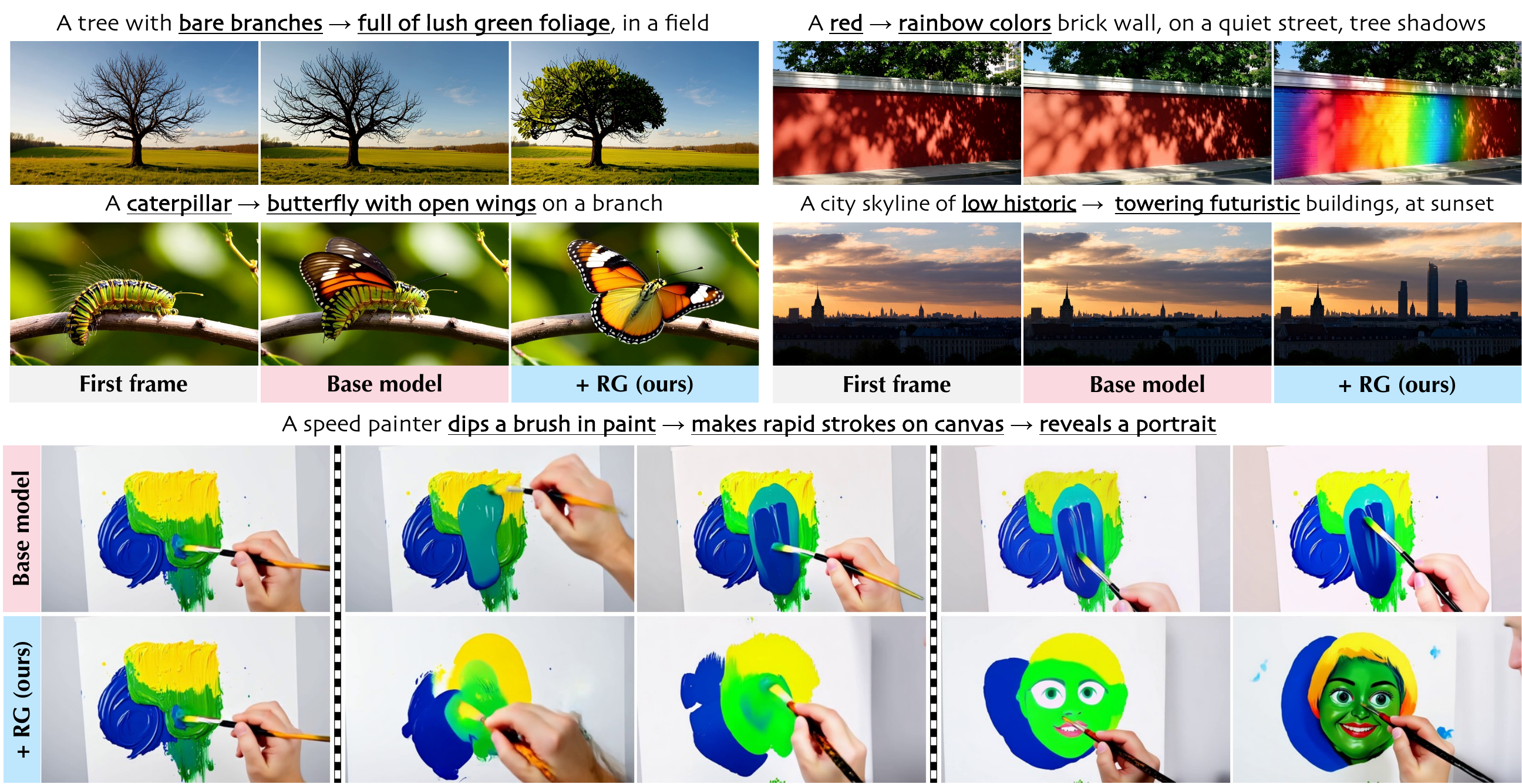}
  \end{center}
  \vspace{-10pt}
  \captionof{figure}{\textbf{Residual guidance (RG) makes streaming video generation follow prompt changes.}
    \textbf{Top:} mid-stream attribute switches. For each prompt we show the first frame and the last frame generated by the base streaming model, with and without RG. 
    The base model stays in the pre-switch state until the end owing to \textit{context inertia}, where each chunk is continued from the previous chunk rather than reflecting the new prompt, whereas the same model with RG realizes the changed attribute while retaining the subject and scene layout.
    \textbf{Bottom:} a three-event story from StoryEval. Dashed lines mark prompt switches. The base model fails to advance beyond the first event, while RG reaches each subsequent event. RG is training-free and easily integrates with existing streaming models and cache-based methods.}
  \label{fig:teaser}
  \vspace{15pt}
}

\makeatother
\begin{document}
\maketitle

\begin{abstract}
Real-time interactive video generation allows users to change the prompt during generation to steer the evolving video.
However, existing streaming models, which generate videos chunk by chunk, often keep following the previous prompt after a switch.
We trace this failure to context inertia, where the previous chunk acts as a strong prior that suppresses the new prompt.
Furthermore, we find that this inertia persists even when the cache retains only the previous chunk, yet discarding the context breaks visual continuity, which limits approaches that edit the cache.
Rather than editing the cache, we propose Residual Guidance (RG), a simple, training-free method that counteracts context inertia with a correction measured outside the cache.
Specifically, RG denoises the re-noised previous chunk once under the new prompt without the cache and adds the resulting residual to the early denoising steps of the next chunk, where its content is largely decided and the new prompt is most suppressed.
Across settings ranging from a single switch to long multi-switch generation, RG substantially improves prompt switching capability on existing streaming models, including those with dedicated switching mechanisms, while largely retaining visual quality.
As it requires no training or per-model tuning and adds only one forward pass per chunk, RG can be easily integrated into existing real-time video generation pipelines.
\end{abstract}
\vspace{-20pt}    
\section{Introduction}
\label{sec:intro}

Conventional video diffusion models~\cite{wan2025wan, kong2024hunyuanvideo, yang2025cogvideox, hacohen2024ltx, lin2024open, zheng2024open, zheng2025open, polyak2024movie, ho2022imagen} typically employ bidirectional attention, allowing video tokens to attend to both past and future frames.
Consequently, they generate a predefined sequence of frames jointly, with the prompt specified before generation and the video becoming available only after the entire sequence has been denoised.
This prevents users from interacting with the generation process as the video unfolds.

\begin{figure*}[ht]
  \centering
  \begin{subfigure}[b]{0.25\textwidth}
    \centering
    \includegraphics[width=\linewidth]{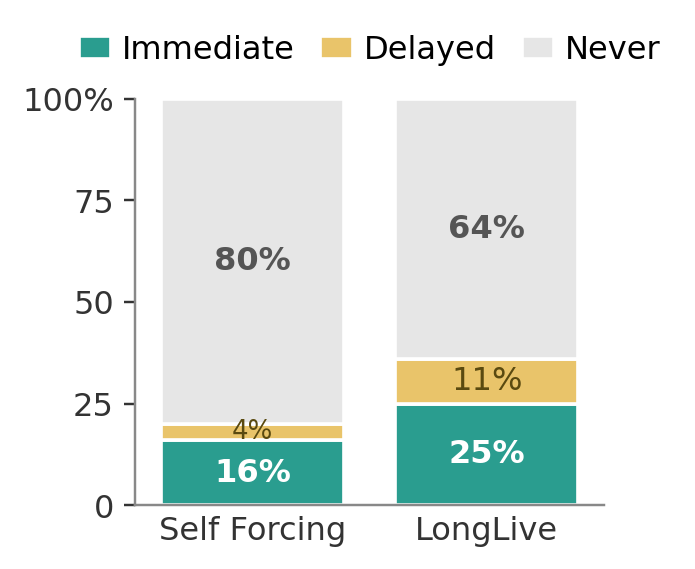}
    \caption{}
    \label{fig:ci_a}
  \end{subfigure}\hfill
  \begin{subfigure}[b]{0.27\textwidth}
    \centering
    \includegraphics[width=\linewidth]{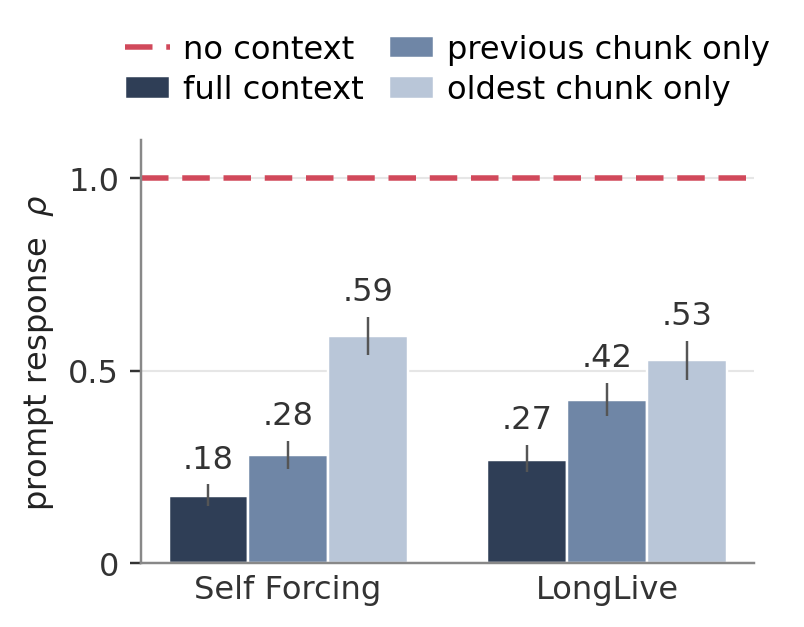}
    \caption{}
    \label{fig:ci_b}
  \end{subfigure}\hspace{5pt}
  \begin{subfigure}[b]{0.46\textwidth}
    \centering
    \includegraphics[width=\linewidth]{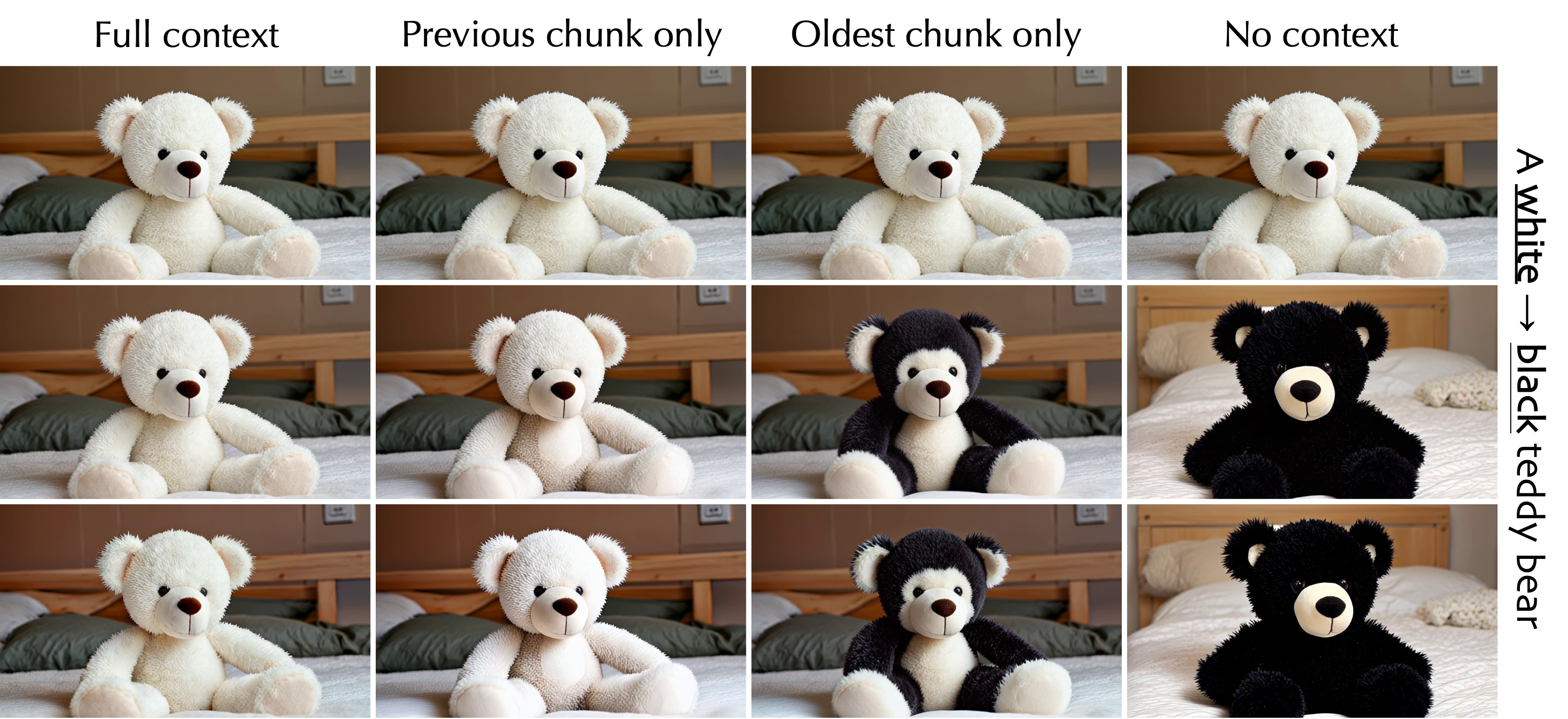}
    \caption{}
    \label{fig:ci_c}
  \end{subfigure}
  \vspace{-5pt}
  \caption{\textbf{Prompt switching under context inertia.}
    (a) Outcomes of 100 prompt switches, most of which are never realized.
    (b) Relative prompt response $\rho$ under different visible contexts. Even a single cached chunk suppresses it.
    (c) Chunks generated under each context (columns) for the same switch. The white bear persists with the full or previous-chunk context, partially darkens with the oldest chunk only, and turns black without context at the cost of identity and background.}
  \label{fig:inertia}
  \vspace{-10pt}
\end{figure*}

Recent autoregressive video diffusion models instead adopt causal attention~\cite{chen2024diffusion}, where each chunk attends to previous chunks through a KV cache~\cite{pope2023efficiently}.
Combined with few-step distillation from a bidirectional teacher~\cite{yin2025slow, huang2026self, yang2025longlive, zhu2026causal, cui2026self, liu2026rolling, kodaira2026streamdit}, such models achieve real-time streaming generation, in which each chunk is displayed as soon as it is generated.
Since the video unfolds while it is being generated, users can change the prompt along the way to steer the evolving video~\cite{yang2025longlive, ji2025memflow, tan2026swift, rao2026no, liu2026advancing, yang2026anchor, yesiltepe2026infinity}, for example, to introduce a new event, modify an existing scene, or control a simulated environment.
However, this interaction remains unreliable.
As shown in Fig.~\ref{fig:teaser} and Fig.~\ref{fig:ci_a}, streaming models often fail to reflect prompt changes and continue to generate content consistent with the previous prompt.
Existing approaches modify the KV cache at prompt boundaries to improve responsiveness~\cite{yang2025longlive, yesiltepe2026infinity}, yet the problem persists.

To understand this failure, we analyze the model's prediction of the switch chunk (Fig.~\ref{fig:ci_b}) and find that the cached visual content acts as a strong prior that suppresses the new prompt, a phenomenon we term \textbf{context inertia}.
Our analysis reveals two properties of context inertia.
First, the inertia comes from what the cached frames depict rather than from the prompt used to generate them, and persists even when only the previous chunk is cached.
This explains why cache-based approaches fall short, as re-caching~\cite{yang2025longlive} updates only the prompt information while keeping the cached frames, and flushing the cache~\cite{yesiltepe2026infinity} still retains the sink and the last frame.
Second, the suppression is most pronounced at the first denoising step, where the content of the chunk is largely determined.
Removing the context entirely would eliminate the inertia but would break subject identity and scene continuity (Fig.~\ref{fig:ci_c}), suggesting that the cache should be kept and its pull countered in the prediction instead.
A natural way to do so is classifier-free guidance~\cite{ho2022classifier}, but its direction is computed under the same cache and is likewise suppressed (see Supp.~\ref{supp:cfg}).
Together, these observations suggest that the prompt signal should be measured without the cache, where context inertia is absent, and applied where the inertia is strongest.

Based on this insight, we propose \textbf{Residual Guidance (RG)}, a simple, training-free method that strengthens the new prompt without cache modification.
Specifically, RG denoises a re-noised copy of the preceding chunk under the new prompt without the cache and adds its difference from the original chunk to the early denoising steps of the next chunk.
This residual captures how the previous content would be redrawn under the new prompt without context inertia, releasing the prediction from that content and supplying the suppressed prompt signal.
Because the correction acts on the prediction rather than on the cache, each chunk still attends to its full context, which preserves subject identity and scene continuity.
For the same reason, RG is orthogonal to how the cache is managed and can be combined with existing cache-based switching methods.
RG requires only one additional forward pass per chunk and uses a single configuration across models.

We evaluate our approach on seven streaming baselines, including methods with dedicated switching mechanisms, across settings ranging from a single switch in a 5\secs video to six prompts over 60\secs.
RG improves average switch achievement from 21\% to 58\% on single-attribute switches, from 47\% to 66\% on three-event stories, and from 66\% to 81\% on long interactive generation, while largely preserving visual quality.
The improvements hold across all baselines and persist over successive switches.
In a user study, RG achieves an overall preference score of 70.7\%.
RG reduces throughput by only 6\% in the single-switch setting, making it practical for real-time generation.

Our contributions are summarized as follows:
(1) We characterize \textit{context inertia} in streaming video generation, showing that cached visual content, rather than stale prompt information, suppresses new prompts, particularly at the first denoising step.
(2) We propose \textit{Residual Guidance}, a simple, training-free method that counteracts context inertia in the prediction rather than in the cache, by measuring the effect of the new prompt without the cache and applying it where the inertia is strongest.
(3) We demonstrate its effectiveness on seven streaming baselines across diverse switching scenarios, with consistent gains in prompt-switch responsiveness at minimal cost in quality and speed.

\section{Related Work}
\label{sec:rwork}

\paragraph{Prompt switching in real-time video generation.}
Recent streaming models distill bidirectional teachers into few-step causal students~\cite{yin2025slow, huang2026self, yang2025longlive, zhu2026causal, cui2026self, liu2026rolling, kodaira2026streamdit} that condition each chunk on previously generated content through a KV cache~\cite{pope2023efficiently}.
To handle prompt switches, LongLive~\cite{yang2025longlive} attributes the failure to stale prompt information in the cache and re-encodes past frames under the new prompt, while Infinity-RoPE~\cite{yesiltepe2026infinity} keeps only a sink and the last frame and Anchor Forcing~\cite{yang2026anchor} rebuilds the cache from stored anchor states.
MemFlow~\cite{ji2025memflow} and SWIFT~\cite{tan2026swift} instead manage the memory by retrieving prompt-relevant history and injecting transition-aware signals, respectively.
Despite their different memory strategies, these methods all retain previously generated visual content in the cache, which we find continues to suppress the new prompt (Sec.~\ref{sec:analysis}). 
RG instead counteracts this suppression in the prediction rather than in the cache, which preserves continuity and keeps it compatible with these approaches.

\vspace{-5pt}
\paragraph{Guidance in video generation.}
Classifier-free guidance (CFG)~\cite{ho2022classifier} strengthens conditioning by extrapolating the conditional prediction away from the unconditional one and is widely used in video diffusion models~\cite{wan2025wan, yang2025cogvideox, hacohen2024ltx}.
Streaming students, however, are distilled at a fixed guidance scale~\cite{yin2025slow}, and applying CFG on top of this distilled guidance only partially restores prompt responsiveness.
Its two branches are computed under the same cache, so the guidance direction is suppressed along with the prompt, and larger scales introduce artifacts (Supp.~\ref{supp:cfg}).
History Guidance~\cite{song2025history} combines predictions across history conditions, enabled by its frame-wise noise-level training, to improve temporal consistency rather than prompt responsiveness.
MAGI-1~\cite{teng2025magi} uses separate history and text guidance scales, but only its non-distilled model exposes an independently adjustable text guidance.
In contrast, RG measures a cache-free prediction with an already generated chunk, so its direction is not suppressed by the context, and it requires no unconditional branch.
A detailed comparison is provided in Supp.~\ref{supp:rwork}.

\vspace{-5pt}
\paragraph{Diffusion-based editing.}
Many diffusion- and flow-based editing methods derive editing signals from source-target prediction differences. Delta Denoising Score~\cite{hertz2023delta} and Contrastive Denoising Score~\cite{nam2024contrastive} use denoising prediction differences, while DreamMotion~\cite{jeong2024dreammotion} uses score distillation with space-time self-similarity constraints. Delta Rectified Flow Sampling~\cite{beaudouin2026delta} extends delta-based guidance to velocity fields. FlowEdit~\cite{kulikov2025flowedit} transports between source and target distributions, and FlowDirector~\cite{li2026flowdirector} uses ODE-based flow corrections. ProxEdit~\cite{han2024proxedit} regularizes source-target differences, while TurboEdit~\cite{deutch2024turboedit} and ChordEdit~\cite{lu2026chordedit} enable few-step and one-step image editing, respectively. StreamEdit~\cite{jiao2026streamedit} also builds on few-step streaming video generators without training, but edits a source video by conditioning a dual-branch sampler on it, whereas RG has no source video and counteracts the cached context of the ongoing generation.
SDEdit~\cite{meng2021sdedit} simply re-noises an input and reconstructs it for image editing. 
RG also re-noises the preceding chunk, but applies the difference between its cache-free reconstruction and the original chunk to the next chunk, without requiring the previous prompt or inversion.
\section{Context Inertia}
\label{sec:analysis}

\paragraph{Notation.}
A streaming model generates video autoregressively in units of one or more latent frames, referred to as chunks.
We denote the clean latent of chunk $k$ by $s_k$ and its noisy state at noise level $t$ by $x_k^t = \alpha_t s_k + \sigma_t \epsilon$.
Chunk $k$ is denoised in a few steps $t_1 > t_2 > \dots > t_Q$ (typically $Q=4$) by a student denoiser $\hat{x}_\theta$, which predicts the clean chunk from its noisy state, conditioned on a prompt $y_k$ and a causal KV cache $\mathcal{C}_{<k}$ storing key--value states computed from preceding chunks $s_1, \dots, s_{k-1}$.
We denote this prediction by $\hat{x}_\theta(x_k^t \mid \mathcal{C}_{<k}, y_k)$, or simply $\hat{x}(\mathcal{C}, y)$ for a fixed noisy input.
A \emph{prompt switch} occurs at chunk $k$ when the prompt changes from $\yold = y_{k-1}$ to $\ynew = y_k \neq \yold$.
We call chunk $k$ the \emph{switch chunk}, so that $s_{k-1}$ is the last chunk generated under $\yold$.

\subsection{Prompt Switches Are Rarely Reflected}
\label{sec:late}
We analyze Self Forcing~\cite{huang2026self}, a representative causal streaming student whose rolling KV cache is retained across the switch, and LongLive~\cite{yang2025longlive}, an interactive model that recomputes cached states under the new prompt at prompt boundaries.
We build a development set of 100 switch pairs $(\yold, \ynew)$ with an LLM~\cite{openai2026gpt55}, covering diverse changes (e.g., color and motion), where $\yold$ and $\ynew$ describe the states before and after the change.
For each pair, we generate a 10\secs video of 14 chunks with the switch at 5\secs, leaving 7 chunks to reflect the new prompt.
A VLM~\cite{bai2025qwen3} then judges whether each post-switch chunk reflects $\ynew$.
We categorize a switch as \emph{immediate} if $\ynew$ appears within the first 2 chunks after the switch (about 1.5\secs, allowing one chunk for a smooth transition), \emph{delayed} if it appears in a later chunk, and \emph{never} if it does not appear within the video.
As shown in Fig.~\ref{fig:ci_a}, only 16\% and 25\% of switches are immediate for Self Forcing and LongLive, respectively, while 80\% and 64\% are never realized.
Supp.~\ref{supp:ci_eval} details the prompt construction and VLM evaluation, and Supp.~\ref{supp:ci_category} reports the results by type of prompt change.

\subsection{Why Do Prompt Switches Fail?}
\label{subsec:cause}
To isolate the effects of the prompt and the context, we run the model unchanged up to the switch chunk $k$ and fix its noisy input $x_k^{t}$ at each denoising step.
From this fixed input, we compute $\hat{x}(\mathcal{C}, y)$ under a visible context $\mathcal{C} \subseteq \mathcal{C}_{<k}$, restricted by hiding cached keys from self-attention, where $\mathcal{C} = \varnothing$ hides all of them.
We measure the prompt response as the displacement induced by switching from $\yold$ to $\ynew$, normalized by the same displacement without context:
\begin{equation}
  \rho(\mathcal{C}) =
  \frac{\|\hat{x}(\mathcal{C}, \ynew) - \hat{x}(\mathcal{C}, \yold)\|}
       {\|\hat{x}(\varnothing, \ynew) - \hat{x}(\varnothing, \yold)\|},
  \label{eq:rho}
\end{equation}
so that $\rho(\mathcal{C})$ is the fraction of the context-free prompt response retained under $\mathcal{C}$, with $\rho(\varnothing) = 1$.
For the numerator, since the cache may carry information about the prompt under which it was built, we recache LongLive under $\ynew$, so that the suppression cannot be attributed to stale prompt information.
For the denominator, $\mathcal{C} = \varnothing$ reproduces the context-free setting of the first chunk of every video, which keeps the evaluation in distribution.

\vspace{-5pt}
\paragraph{Finding 1: The context suppresses the prompt.}
We examine how the cached context affects the response to the new prompt, both quantitatively and qualitatively.
Unless otherwise noted, we report $\rho$ at the first denoising step, which we show below to be the most affected.
As shown in Fig.~\ref{fig:ci_b}, at the first denoising step the full context retains only 0.18 and 0.27 of the context-free prompt response for Self Forcing and LongLive, respectively, a suppression of the new prompt by cached content that we refer to as \emph{context inertia}.
Furthermore, with the previous chunk alone, $\rho$ reaches only 0.28 and 0.42 for the two models, and even with the oldest chunk alone it stays at 0.59 and 0.53, far below 1.
A single recent chunk thus accounts for most of the suppression, and even distant frames contribute.
Since LongLive is measured after recaching, this suppression stems largely from the cached visual content itself rather than stale prompt information.
The generated chunks in Fig.~\ref{fig:ci_c} exhibit the same trend: the white bear persists under the full context or the previous chunk alone and only partially darkens with the oldest chunk.
Removing the context entirely finally produces the black bear, but at the cost of its identity and background, since the chunk is then determined solely by the noise and the prompt.
One might expect stronger classifier-free guidance to counteract this suppression.
Since the guidance direction $\hat{x}(\mathcal{C}, \ynew) - \hat{x}(\mathcal{C}, y_\varnothing)$, where $y_\varnothing$ denotes the empty prompt, is computed under the same cache, it is suppressed as well, and a larger guidance scale mainly amplifies artifacts rather than restoring the prompt response (Supp.~\ref{supp:cfg}).

\vspace{-5pt}
\paragraph{Finding 2: Context inertia is strongest at the first step.}
To see where along the few-step trajectory this suppression arises, we repeat the measurement at every denoising step and find that two distinct effects both peak at the first step: the prompt alone moves the prediction most, and the context suppresses this movement most.
First, without context, the change in the clean prediction of the switch chunk induced by switching from $\yold$ to $\ynew$, $\|\hat{x}(\varnothing, \ynew) - \hat{x}(\varnothing, \yold)\| / \|s_{k-1}\|$, is 0.62 at step 1 for Self Forcing, but only 0.19 at step 2 and 0.03 at step 4.
Second, with context, the fraction of this response that survives, $\rho$, is lowest at step 1, and the response lost there is not recovered at later steps (see Supp.~\ref{supp:ci_step} for details).
We attribute both to the coarse-to-fine nature of diffusion sampling, where early high-noise steps largely determine layout and content while later steps refine details~\cite{choi2022perception, hertz2022prompt, balaji2022ediff}, an effect likely amplified in few-step students whose first step covers most of the noise range.
\section{Method}
\label{sec:method}

Sec.~\ref{sec:analysis} suggests that following the new prompt requires evaluating it \emph{without} the cache, whereas preserving visual continuity requires generating \emph{with} it.
\emph{Residual Guidance (RG)} satisfies both by measuring the prompt signal without the cache while generating the chunk with it.
Specifically, RG reconstructs a re-noised copy of the preceding chunk under the new prompt without the cache, and adds its residual from the original chunk to the cache-conditioned prediction of the next chunk at the early denoising steps, where context inertia is strongest.

\subsection{A Posterior View of the Problem}
The intuition above has a simple probabilistic interpretation.
The clean prediction of the student approximates the posterior mean of $s_k$ given its noisy state $x_k^t$, the preceding cache $\mathcal{C}_{<k}$, and the prompt $\ynew$.
Since the forward noising process does not depend on the prompt (\ie, $x_k^t \perp \ynew \mid s_k, \mathcal{C}_{<k}$), this posterior factorizes as
\begin{equation}
p(s_k \mid x_k^t, \mathcal{C}_{<k}, \ynew) \;\propto\; p(s_k \mid x_k^t, \mathcal{C}_{<k})\, p(\ynew \mid s_k, \mathcal{C}_{<k}).
\label{eq:posterior}
\end{equation}
The first term favors continuing the cached content, and the second favors the new prompt.
Context inertia arises when the first term dominates the second.
Methods that edit the cache at a switch~\cite{yang2025longlive, yesiltepe2026infinity} weaken the first term, and stronger text guidance amplifies the second, yet this term is still computed under the cache, so much of the suppression remains.
RG instead adds a residual measured without the cache, which pushes the prediction toward the new prompt and away from the cached content. Both effects contribute to the gain (Tab.~\ref{tab:alternatives}).
At the same time, since the chunk is still generated under the full cache, visual continuity with the preceding chunks is largely retained.

\subsection{Residual Guidance}
\label{sec:rg}

\paragraph{Residual estimation.}
To measure the effect of the new prompt without context inertia, the preceding chunk $s_{k-1}$ is first re-noised to a noise level $\tau$ and then denoised in a single step under $\ynew$ without the KV cache:
\begin{align}
\tilde{x}^{\tau}_{k-1} &= \alpha_\tau s_{k-1} + \sigma_\tau \epsilon, \qquad \epsilon \sim \mathcal{N}(0, I), \nonumber \\
r_k &= \hat{x}_\theta(\tilde{x}^{\tau}_{k-1} \mid \varnothing, \ynew) - s_{k-1}.
\label{eq:residual}
\end{align}
Here, $\varnothing$ denotes an empty KV cache, so the preceding chunk enters the reconstruction only through its re-noised latent.
The residual measures how the cache-free reconstruction under $\ynew$ differs from the original chunk, and part of it is independent of $\ynew$, counteracting the previous content itself.
It is thus not an exact estimate of the prompt term in Eq.~\eqref{eq:posterior} but an approximate correction to the cache-conditioned prediction (see Supp.~\ref{supp:theory}).
The noise level $\tau$, chosen independently of the sampling steps $t_i$, sets how much of the preceding chunk the reconstruction retains and thus the strength of the correction, analogous to the guidance scale in classifier-free guidance~\cite{ho2022classifier}.

\vspace{-5pt}
\paragraph{Residual injection.}
The residual $r_k$ is measured on chunk $k-1$ and applied to chunk $k$.
Context inertia itself motivates this transfer. Since the cache-conditioned prediction of chunk $k$ stays close to $s_{k-1}$, a correction derived from $s_{k-1}$ also applies to it.
The residual is added to the clean prediction at the early denoising steps:
\begin{equation}
\hat{x}_k^{(i)} = \hat{x}_\theta(x_k^{t_i} \mid \mathcal{C}_{<k}, \ynew) + \lambda_i r_k,
\label{eq:inject}
\end{equation}
with $\lambda = (1, 0.5, 0, 0)$ over the four steps.
The correction thus acts where the suppression is strongest (Sec.~\ref{sec:analysis}), while the remaining steps run unmodified under the full cache and keep the chunk consistent with its history.
Unlike classifier-free guidance, which contrasts two predictions with the cache, the residual is measured without the cache and is therefore not attenuated by it.

\vspace{-5pt}
\paragraph{Implementation.} 
For every chunk generated under $\ynew$, the residual is recomputed from its preceding chunk, until the next prompt switch.
We set $\tau = 850$ for all models and sample $\epsilon$ once per switch, sharing it across the prompt interval (Supp.~\ref{supp:ablation}).
Alternative interventions, including residuals against the empty or previous prompt, partial cache removal, and stronger classifier-free guidance, yield smaller gains or degrade visual continuity (Supp.~\ref{supp:alternatives}).
Both ablations and alternatives are examined in Sec.~\ref{sec:ablation}, and the full algorithm is given in Supp.~\ref{supp:algorithm}.
\section{Experiments}
\label{sec:exp}

\begin{table*}[t]
\begin{minipage}[t]{0.645\textwidth}
\centering
\caption{\textbf{Prompt switching on T2V-CompBench dynamic attribute.}
Achievement: fraction of videos reflecting the new prompt.
T2V-CB: official benchmark score~\cite{sun2025t2v}.
Quality and Overall follow VBench~\cite{huang2023vbench}, with Achievement as the semantic score.}
\label{tab:dynattr}
\vspace{-5pt}
\footnotesize
\setlength{\tabcolsep}{3pt}
\renewcommand{\arraystretch}{0.9}
\resizebox{\linewidth}{!}{%
\begin{tabular}{l cccccc}
\toprule
\multirow{2}{*}{\textbf{Method}}
& \multirow{2}{*}{\textbf{Achievement $\uparrow$}}
& \textbf{T2V-CB}
& \multirow{2}{*}{\textbf{CLIP $\uparrow$}}
& \textbf{Dynamic}
& \multirow{2}{*}{\textbf{Quality $\uparrow$}}
& \multirow{2}{*}{\textbf{Overall $\uparrow$}} \\
&
& \textbf{Dyn.\ Attr.\ $\uparrow$}
&
& \textbf{Degree $\uparrow$}
& & \\
\midrule

Self Forcing
& 0.106 & 0.013 & 24.97 & 0.250 & \underline{0.813} & 0.672 \\
\oursrow
+\,\ours
& 0.568 & 0.119 & 26.29 & 0.380 & 0.806 & 0.758 \\

Causal Forcing
& 0.147 & 0.024 & 24.78 & 0.425 & \textbf{0.821} & 0.686 \\
\oursrow
+\,\ours
& 0.558 & 0.085 & 25.68 & \textbf{0.600} & \underline{0.813} & 0.762 \\

MemFlow
& 0.210 & 0.035 & 25.55 & 0.175 & 0.805 & 0.686 \\
\oursrow
+\,\ours
& 0.493 & 0.091 & 26.41 & 0.235 & 0.802 & 0.740 \\

\midrule
LongLive
& 0.296 & 0.039 & 25.46 & 0.145 & 0.805 & 0.703 \\
\oursrow
+\,\ours
& \underline{0.622} & \underline{0.145} & \underline{26.82} & 0.215 & 0.802 & 0.766 \\

Infinity-RoPE
& 0.148 & 0.023 & 24.75 & 0.155 & 0.806 & 0.674 \\
\oursrow
+\,\ours
& 0.578 & 0.110 & 25.99 & 0.310 & 0.803 & 0.758 \\

\midrule
Anchor Forcing
& 0.328 & 0.058 & 25.61 & 0.365 & 0.811 & 0.714 \\
\oursrow
+\,\ours
& 0.606 & \textbf{0.152} & 26.45 & \underline{0.540} & \underline{0.813} & \underline{0.772} \\

SWIFT
& 0.252 & 0.028 & 25.32 & 0.120 & 0.806 & 0.695 \\
\oursrow
+\,\ours
& \textbf{0.667} & 0.144 & \textbf{26.86} & 0.150 & 0.799 & \textbf{0.773} \\
\bottomrule
\end{tabular}%
}
\end{minipage}%
\hfill
\begin{minipage}[t]{0.34\textwidth}
\centering
\vspace{-5pt}
\includegraphics[width=\linewidth]{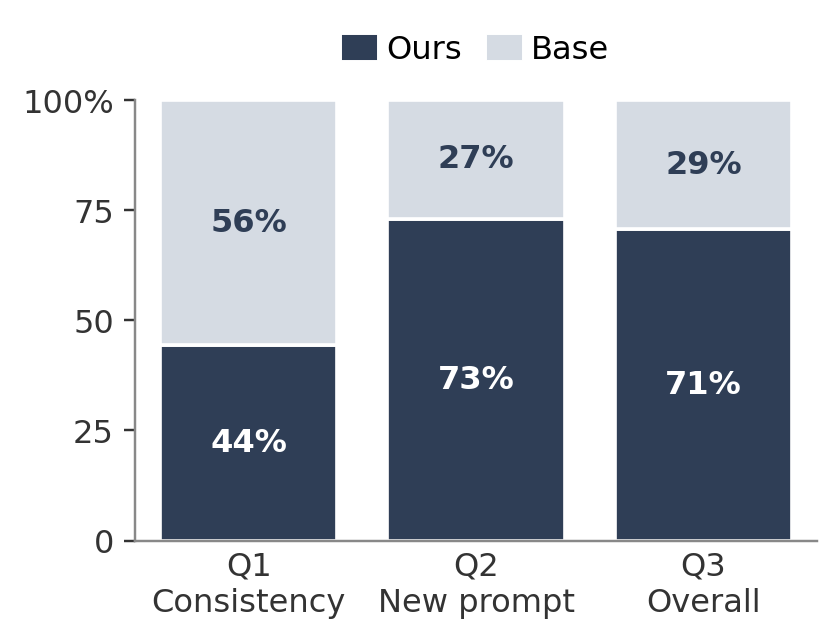}
\vspace{-20pt}
\captionof{figure}{\textbf{User study.}
Preference (\%) for \emph{+\,\ours} on the same prompt pair from 20 raters, judged on consistency (no unwanted change beyond the requested one), adherence to the new prompt, and overall preference.
RG is largely preferred for adherence and overall, with consistency slightly below parity.}
\label{fig:userstudy}
\end{minipage}
\vspace{-10pt}
\end{table*}

\begin{figure*}[ht]
\centering
\includegraphics[width=\textwidth]{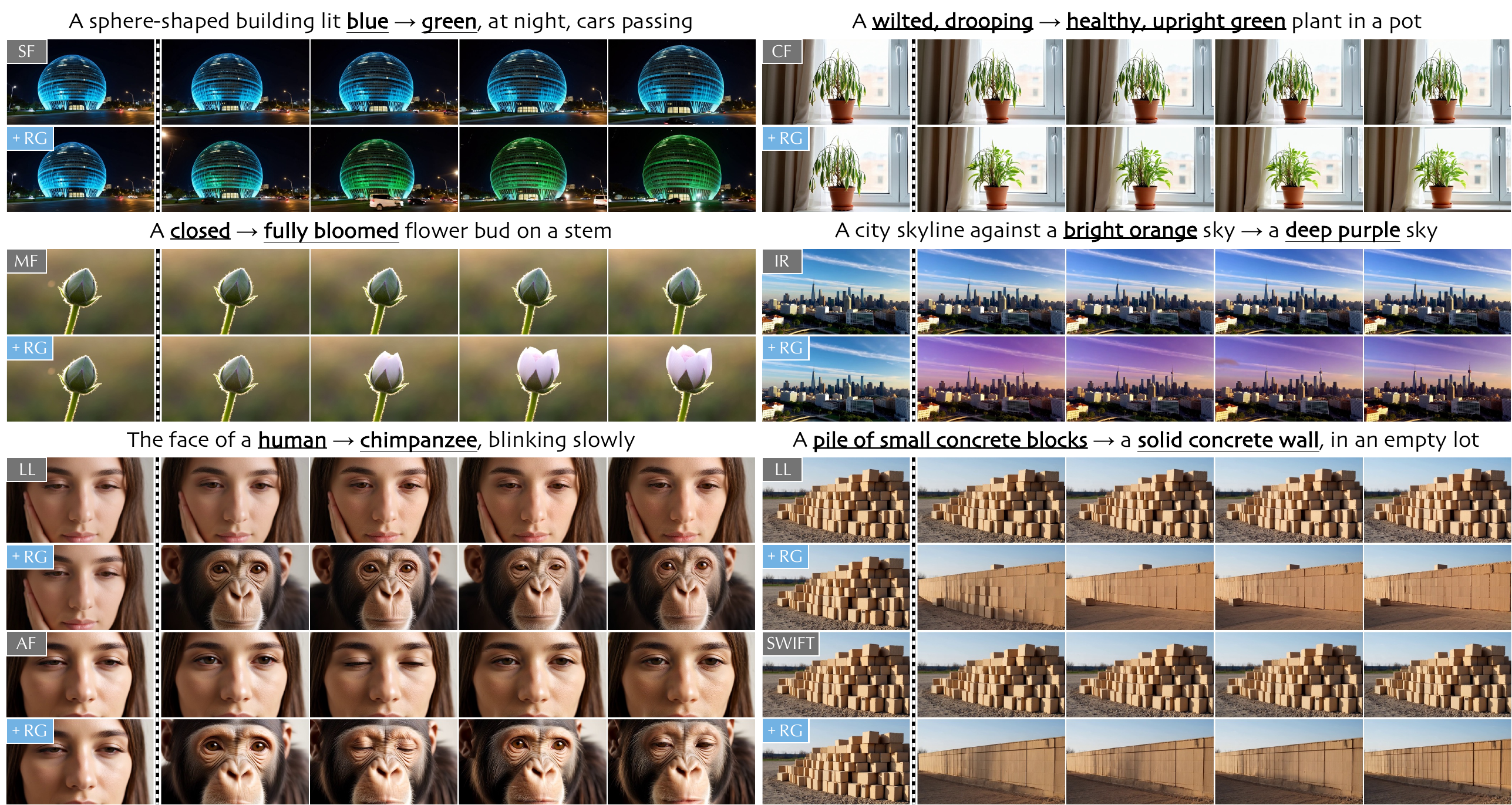}
\vspace{-15pt}
\caption{\textbf{Qualitative results on T2V-CompBench.}
The white-to-black bar marks the prompt switch.
Without RG, the video keeps the initial state after the switch, whereas RG enables the new state to appear while the subject and scene layout remain consistent.
SF: Self Forcing, CF: Causal Forcing, MF: MemFlow, IR: Infinity-RoPE, LL: LongLive, AF: Anchor Forcing.
}
\label{fig:quali_dynattr}
\vspace{-10pt}
\end{figure*}
\begin{table*}[h!]
\centering
\small
\caption{\textbf{Multi-prompt and longer settings.}
StoryEval: 100 three-event prompts over 10\secs with switches. 
InteractiveBench: 100 scripts of six 10\secs prompts over 60\secs, following MemFlow~\cite{ji2025memflow} and LongLive~\cite{yang2025longlive}. Rows marked
\emph{+ \ours} apply residual guidance to the model above.}
\vspace{-5pt}
\label{tab:longer}
\setlength{\tabcolsep}{4pt}
\resizebox{\textwidth}{!}{%
\begin{tabular}{l ccccc @{\hskip 12pt} c ccccc c c c}
\toprule
& \multicolumn{5}{c}{\textbf{StoryEval (10\secs)}}
& \multicolumn{9}{c}{\textbf{InteractiveBench (60\secs)}} \\
\cmidrule(lr){2-6}
\cmidrule(lr){7-15}
& \multirow{2}{*}{\textbf{Achv.\ $\uparrow$}}
& \multirow{2}{*}{\textbf{CLIP $\uparrow$}}
& \textbf{Dynamic}
& \multirow{2}{*}{\textbf{Quality $\uparrow$}}
& \multirow{2}{*}{\textbf{Overall $\uparrow$}}
& \multirow{2}{*}{\textbf{Achv.\ $\uparrow$}}
& \multicolumn{5}{c}{\textbf{CLIP $\uparrow$}}
& \textbf{Dynamic}
& \multirow{2}{*}{\textbf{Quality $\uparrow$}}
& \multirow{2}{*}{\textbf{Overall $\uparrow$}} \\
\cmidrule(lr){8-12}
\textbf{Method} & & & \textbf{Degree $\uparrow$} & & & & 10--20 & 20--30 & 30--40 & 40--50 & 50--60 & \textbf{Degree $\uparrow$} & & \\
\midrule
MemFlow
& 0.385 & 22.41 & 0.625 & \textbf{0.830} & 0.741
& 0.587 & 27.04 & 27.38 & 26.98 & 26.89 & 26.38 & 0.254 & 0.810 & 0.766 \\
\oursrow + \ours
& 0.659 & 23.01 & 0.750 & \underline{0.826} & \underline{0.793}
& 0.724 & 27.79 & 27.92 & 27.49 & 27.73 & 27.39 & 0.302 & 0.806 & 0.790 \\
\midrule
LongLive
& 0.511 & 22.31 & 0.515 & 0.824 & 0.762
& 0.631 & 27.35 & 27.73 & 27.58 & 27.78 & 26.81 & 0.284 & 0.815 & 0.778 \\
\oursrow + \ours
& \underline{0.687} & \textbf{23.13} & 0.680 & 0.823 & \textbf{0.795}
& \underline{0.826} & \textbf{28.03} & \textbf{28.38} & \underline{27.73} & \textbf{28.19} & \textbf{27.49} & 0.340 & 0.809 & 0.812 \\
Infinity-RoPE
& 0.402 & 21.87 & 0.530 & 0.821 & 0.738
& 0.609 & 26.83 & 27.19 & 26.85 & 26.65 & 26.30 & 0.388 & 0.824 & 0.781 \\
\oursrow + \ours
& 0.586 & 22.58 & 0.765 & 0.819 & 0.772
& 0.783 & 27.10 & 27.55 & 27.16 & 27.03 & 27.07 & 0.560 & 0.827 & \underline{0.818} \\
\midrule
Anchor Forcing
& 0.538 & 22.59 & \underline{0.895} & 0.821 & 0.764
& 0.765 & 26.98 & 27.12 & 27.03 & 26.75 & 26.62 & \underline{0.796} & \underline{0.831} & \underline{0.818} \\
\oursrow + \ours
& \textbf{0.732} & \underline{23.11} & \textbf{0.965} & 0.802 & 0.788
& \textbf{0.871} & 27.32 & 27.54 & 27.41 & 27.25 & \underline{27.43} & \textbf{0.886} & \textbf{0.832} & \textbf{0.840} \\
SWIFT
& 0.505 & 22.23 & 0.465 & 0.823 & 0.759
& 0.693 & 27.54 & 27.90 & 27.51 & 27.71 & 26.84 & 0.306 & 0.815 & 0.791 \\
\oursrow + \ours
& 0.654 & 23.08 & 0.645 & 0.821 & 0.788
& 0.823 & \underline{27.86} & \underline{28.17} & \textbf{27.82} & \underline{27.82} & 27.42 & 0.394 & 0.811 & 0.813 \\
\bottomrule
\end{tabular}%
}
\vspace{-10pt}
\end{table*}

\begin{figure*}[h!]
    \centering
    \begin{subfigure}{\linewidth}
        \centering
        \includegraphics[width=\linewidth]{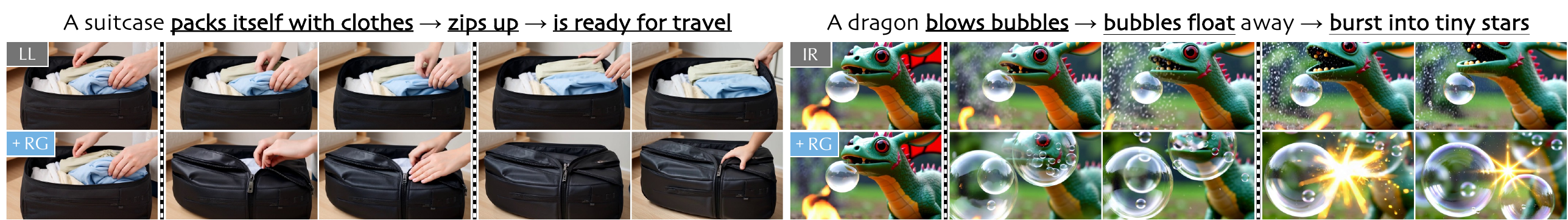}
        \caption{StoryEval}
        \label{fig:quali_storyeval}
    \end{subfigure}
    \vspace{5pt}
    \begin{subfigure}{\linewidth}
        \centering
        \includegraphics[width=\linewidth]{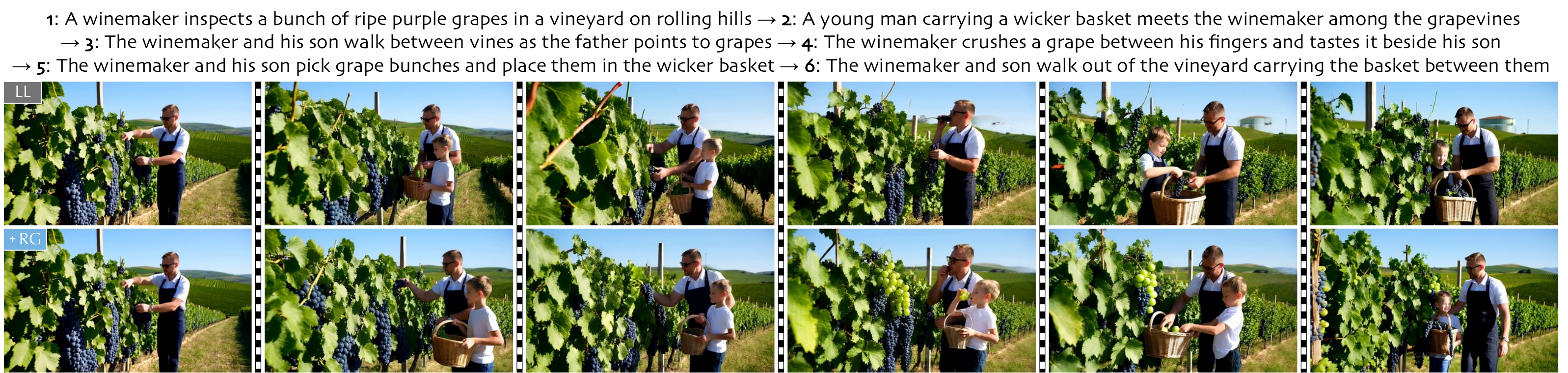}
        \caption{InteractiveBench}
        \label{fig:quali_interactive}
    \end{subfigure}
    \vspace{-20pt}
    \caption{\textbf{Qualitative results on multi-event generation.} In each example, the top row is the base model and the bottom row is the same model with RG. Dashed lines mark prompt switches. (a) Three-event stories (10\secs). The base models remain in the first event throughout, whereas RG advances to each event as it is prompted. (b) A 60\secs interactive sequence with six prompts. RG follows each instruction in turn, including the appearance and interaction of a new subject.}
    \label{fig:quali_multievent}
    \vspace{-12pt}
\end{figure*}

\subsection{Setup}

\paragraph{Baselines.}
We apply RG to Self Forcing~\cite{huang2026self}, Causal Forcing~\cite{zhu2026causal}, and MemFlow~\cite{ji2025memflow}, which retain the cache at switches, to LongLive~\cite{yang2025longlive} and Infinity-RoPE~\cite{yesiltepe2026infinity}, which modify the cache, and to Anchor Forcing~\cite{yang2026anchor} and SWIFT~\cite{tan2026swift}, which use dedicated switch mechanisms.
Since RG leaves the cache untouched, it plugs into all seven baselines regardless of how they manage the cache, without retraining or per-model tuning, using the same configuration throughout.
All are distilled from Wan2.1-1.3B~\cite{wan2025wan}, generate $832\times480$ videos, and use their native sampling.
Each model and its RG variant share the same seed, ensuring identical pre-switch frames.
We further apply our method to Rolling Forcing~\cite{liu2026rolling} in Supp.~\ref{supp:rolling}.

\paragraph{Datasets.}
We evaluate RG in three settings of increasing duration and number of switches, each probing a different aspect of prompt switching.
First, the 200 dynamic attribute prompts of T2V-CompBench~\cite{sun2025t2v} are split into initial and final states and generated as 5\secs videos with the switch at 2\secs.
Since only one attribute of the same subject changes, this setting most directly tests context inertia.
Second, 100 three-event stories from StoryEval~\cite{wang2024storyeval} are generated as 10\secs videos, with their events used as consecutive prompts and switches at 3\secs and 7\secs, testing whether a model can switch repeatedly.
Third, following MemFlow~\cite{ji2025memflow} and other interactive methods~\cite{yang2025longlive, yang2026anchor, tan2026swift}, we use 100 scripts of six prompts, each active for 10\secs over a 60\secs video, which we refer to as InteractiveBench.
Prompt construction is detailed in Supp.~\ref{supp:benchmarks}.
Supp.~\ref{supp:actbind}, \ref{supp:240s}, and~\ref{supp:scene_cuts} further cover action-binding prompts, where a new subject appears at the switch, InteractiveBench extended to 240\secs, and explicit scene cuts with Infinity-RoPE~\cite{yesiltepe2026infinity} and its RoPE Cut.

\paragraph{Metrics.}
A VLM judge~\cite{bai2025qwen3} determines whether the post-switch video depicts the new prompt (reliability in Supp.~\ref{supp:judge_reliability}).
\emph{Achievement} is the fraction of switches after which the video depicts the new prompt.
We exclude new prompts that the model cannot render even as an initial prompt, so that failures reflect the switch rather than the prompt.
On the post-switch segment, we report CLIP similarity~\cite{radford2021learning} to the new prompt and the VBench~\cite{huang2023vbench} quality score, which aggregates seven quality dimensions with the official normalization and weights.
We also report VBench dynamic degree, which ensures that consistency is not obtained simply by producing static videos.
Following the VBench total score, \emph{Overall} is computed as $(4\cdot\text{Quality} + \text{Achievement})/5$, with Achievement in place of the semantic score.
For dynamic attributes, we additionally report the official T2V-CompBench score~\cite{sun2025t2v}.
Metric definitions are detailed in Supp.~\ref{supp:metrics}.

\subsection{Main Results}
\paragraph{Single switch.}
Tab.~\ref{tab:dynattr} shows that context inertia severely limits existing models, especially when only a single attribute is requested to change.
Without RG, Achievement is at most 0.21 on Self Forcing, Causal Forcing, MemFlow, and Infinity-RoPE, and 0.25 to 0.33 on LongLive, Anchor Forcing, and SWIFT.
RG raises Achievement by 0.28 to 0.46 on every base model, for example from 0.106 to 0.568 on Self Forcing.
These gains are not specific to the VLM judge, as CLIP and the official T2V-CompBench score, which evaluates both states of the prompt, improve on every base model as well.
The highest scores are obtained when RG is applied on top of LongLive, Anchor Forcing, and SWIFT, indicating that RG complements existing switching mechanisms rather than replacing them.
This improvement comes at little cost, as Quality stays within 0.01 of the base model.
Its slight decrease on most models reflects the known trade-off between consistency metrics and dynamic degree~\cite{huang2023vbench}, which rises on all models as the attribute change itself introduces motion (further discussed in Supp.~\ref{supp:quality}).
Overall therefore improves on every base model, with the two best scores on SWIFT and Anchor Forcing with RG.
Fig.~\ref{fig:quali_dynattr} shows that, with RG, the requested attribute appears promptly after the switch (e.g., the bud blooms, a human face turns into a chimpanzee) while the subject and layout are largely retained (see Supp.~\ref{supp:switch_chunk}), whereas the base models mostly remain in the initial state.

\begin{figure}[t!]
  \centering
  \includegraphics[width=\linewidth]{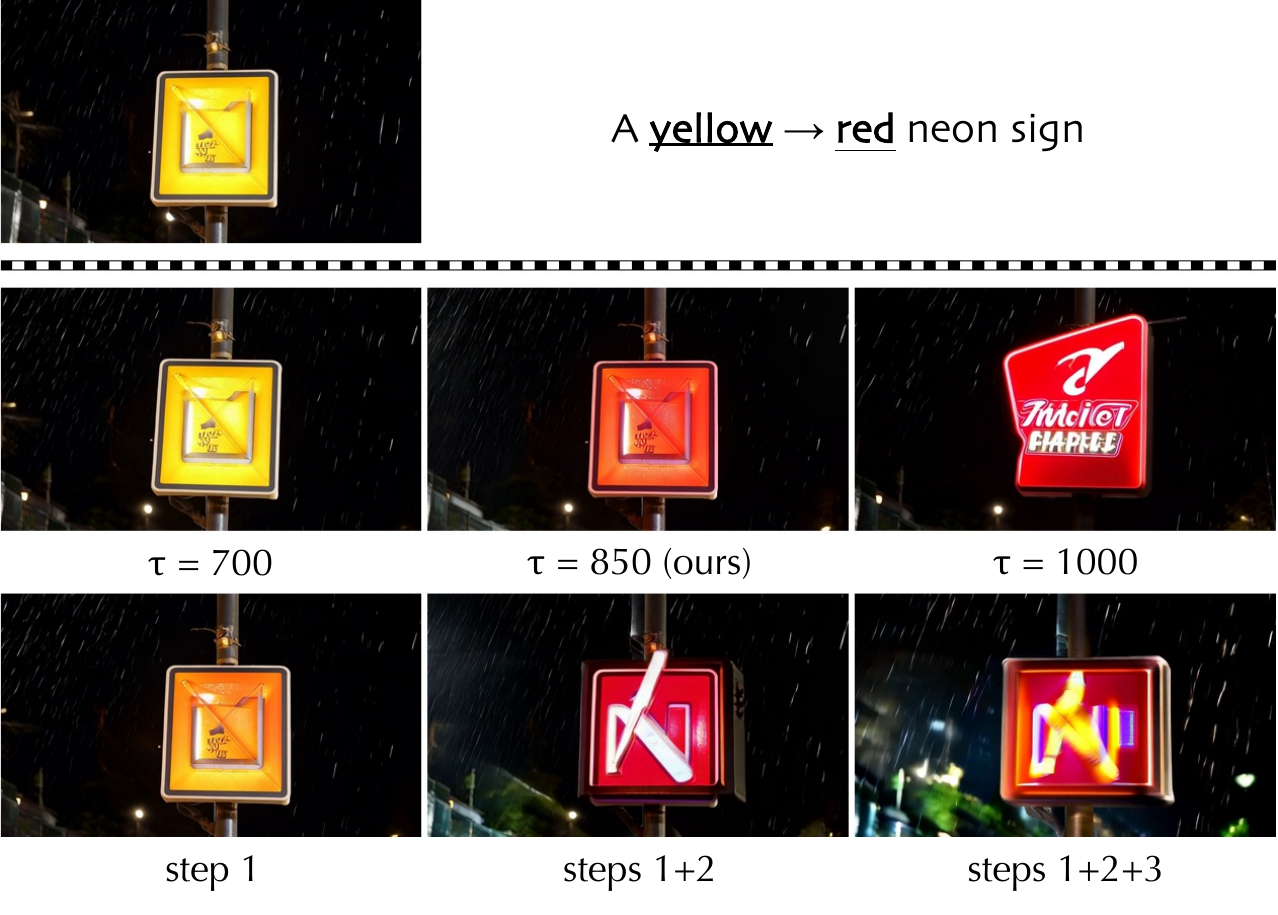}
  \vspace{-22pt}
  \caption{\textbf{Noise level and step schedule.} Top: noise level $\tau$ with our schedule $\lambda = (1, 0.5, 0, 0)$. Bottom: residual added in full ($\lambda_i = 1$) at the listed steps, with $\tau = 850$. The frame before the switch (yellow neon sign) is shown on the top.}
  \label{fig:ablation_qual}
  \vspace{-15pt}
\end{figure}

\vspace{-10pt}
\paragraph{Multiple switches.}
Tab.~\ref{tab:longer} (left) evaluates short multi-switch generation on the three-event stories of StoryEval.
Base Achievement is higher here than on dynamic attributes, since changes in action or scene are rendered more readily than an attribute change on the same subject.
RG still raises Achievement by 0.15 to 0.27 on every base model, e.g., from 0.385 to 0.659 on MemFlow, and improves CLIP and dynamic degree on all models while keeping Quality within 0.02 of the base model.
Overall thus improves throughout and is highest for LongLive and MemFlow with RG.
Fig.~\ref{fig:quali_storyeval} shows that RG realizes both switches in turn (e.g., the suitcase zips up, and the bubbles burst into tiny stars), whereas the base models hardly move beyond the first event.

\vspace{-10pt}
\paragraph{Long interactive generation.}
Tab.~\ref{tab:longer} (right) evaluates long multi-switch generation on InteractiveBench, which follows the 60\secs protocol of prior interactive methods~\cite{ji2025memflow, yang2025longlive}, so that they are tested under their own setting.
RG raises Achievement by 0.11 to 0.20 on every base model, e.g., from 0.631 to 0.826 on LongLive.
The gain is not limited to early switches, as CLIP improves in every 10\secs segment after the first switch, including the last one.
Quality stays within 0.006 of the base model while dynamic degree increases, so Overall improves on all models and is highest for Anchor Forcing with RG.
Fig.~\ref{fig:quali_interactive} shows that LongLive follows the switches only in part (e.g., the winemaker holds the grapes but the son is absent), whereas RG satisfies each prompt well.
A dimension-wise breakdown of VBench-Long and additional qualitative results are provided in Supp.~\ref{supp:vbench_long} and~\ref{supp:qualitative}.

\vspace{-10pt}
\paragraph{User study.}
To assess how users perceive these gains, 20 raters compare each base model with and without RG on the same prompt pairs (Fig.~\ref{fig:userstudy}).
Averaged over base models, RG is preferred for following the new prompt in 73.1\% of comparisons, while consistency remains slightly below parity at 44.3\%.
Asked which result they would prefer when switching prompts in real time, raters choose RG 70.7\% of the time, indicating that the gain in responsiveness outweighs the small loss in consistency in practical use (Supp.~\ref{supp:user_study}).

\vspace{-10pt}
\paragraph{Computational overhead.}
RG adds one cache-free forward pass to each post-switch chunk.
On a single RTX A6000, RG lowers throughput by about 6\% over a 5\,s video with one switch.
Meanwhile, switches take effect faster, with the fraction realized within two chunks rising from 0.21 to 0.54 on LongLive. 
See Supp.~\ref{supp:speed} for details and Supp.~\ref{supp:switch_chunk} for the switch chunk itself.

\subsection{Ablations \& Alternatives}
\label{sec:ablation}

\paragraph{Ablations.}
Fig.~\ref{fig:ablation_qual} visualizes the two main design choices.
The noise level $\tau$ controls how much of the preceding chunk the reconstruction retains, acting as a strength knob analogous to the guidance scale: at $\tau = 700$, the sign remains yellow, at $\tau = 1000$, it is redrawn with a new shape as in the context-free prediction (Fig.~\ref{fig:ci_c}), and at $\tau = 850$, its color changes while its shape is preserved.
The step schedule balances the correction against refinement.
Adding the residual only at step 1 yields a partial switch, as the next step pulls the chunk back toward the cached context.
Adding it in full at step 2 as well, $\lambda = (1, 1, 0, 0)$, counteracts this pull but overrides the shape inherited from the context, and extending it to step 3 produces artifacts.
Halving the weight at step 2, $\lambda = (1, 0.5, 0, 0)$, prevents the pull-back while letting the cache restore the shape, leaving the later steps for refinement.
Quantitative results are provided in Supp.~\ref{supp:ablation}.

\vspace{-5pt}
\paragraph{Alternatives.}
\begin{table}[t]
\centering
\caption{\textbf{Alternatives to residual guidance.} Evaluated on 100 random dynamic attribute pairs with Self Forcing and LongLive. Unint.: unintended change, measured as the DreamSim~\cite{fu2023dreamsim} distance across the switch with the changed object masked out.}
\vspace{-5pt}
\label{tab:alternatives}
\footnotesize
\setlength{\tabcolsep}{3pt}
\renewcommand{\arraystretch}{0.9}
\resizebox{\linewidth}{!}{%
\begin{tabular}{@{}l ccc ccc@{}}
\toprule
 & \multicolumn{3}{c}{Self Forcing} & \multicolumn{3}{c}{LongLive} \\
\cmidrule(lr){2-4}\cmidrule(l){5-7}
Method & Achv.$\uparrow$ & Quality$\uparrow$ & Unint.$\downarrow$ & Achv.$\uparrow$ & Quality$\uparrow$ & Unint.$\downarrow$ \\
\midrule
Base model          & 0.130 & 0.789 & 0.026 & 0.350 & 0.770 & 0.021 \\
\midrule
No context (all)    & 1.000 & 0.726 & 0.191 & 0.963 & 0.733 & 0.163 \\
No context (step 1) & 0.948 & 0.803 & 0.165 & 0.863 & 0.787 & 0.114 \\
CFG ($w{=}3$)       & 0.234 & 0.793 & 0.033 & 0.550 & 0.772 & 0.033 \\
Text-only residual  & 0.351 & 0.788 & 0.037 & 0.525 & 0.772 & 0.032 \\
\midrule
\rowcolor{gray!15}
RG (ours)           & 0.623 & 0.788 & 0.089 & 0.675 & 0.778 & 0.046 \\
\bottomrule
\end{tabular}%
}
\vspace{-12pt}
\end{table}
Tab.~\ref{tab:alternatives} compares RG with simpler interventions.
Removing the context, even at the first step only, realizes most switches but causes 1.9 to 3.5 times as much unintended change as RG.
CFG preserves visual continuity but stays below RG in Achievement, as its prompt direction is also suppressed by the context, while requiring an extra forward pass at every guided step.
The text-only residual, which drops the prompt-independent part $\hat{x}_\theta(\tilde{x}^{\tau}_{k-1}\mid\varnothing,y_\varnothing)-s_{k-1}$ from $r_k$, recovers only about half of the gain, showing that erasing what the context imposes matters as much as strengthening the prompt.
More alternatives are discussed in Supp.~\ref{supp:alternatives}.

\section{Conclusion}
\label{sec:discussion}

In this work, we trace the failure of prompt switching in streaming video generation to \emph{context inertia}: cached visual content suppresses the influence of a new prompt, and existing cache-based switching strategies do not fully remove this suppression.
\emph{Residual Guidance (RG)} addresses the problem by measuring the effect of the new prompt without the cache and adding the resulting residual to the next chunk at its early denoising steps, where context inertia is strongest.
RG needs no training, combines with existing switching methods, and adds only one forward pass per chunk, making it practical for real-time use.
Across seven streaming baselines and settings ranging from a single switch to multiple switches in long video generation, RG substantially improves switch achievement while retaining visual continuity.
Nevertheless, some changes remain difficult, such as removing an object, and fine details can be altered along with the requested change.
We discuss these failure cases in Supp.~\ref{supp:limitations} as directions for future work.

\clearpage

{
    \small
    \bibliographystyle{ieeenat_fullname}
    \bibliography{main}
}

\clearpage
\setcounter{page}{1}
\maketitlesupplementary
\appendix

\section{Analysis Details}
\label{supp:analysis}

\subsection{Context Inertia Prompt Set and Evaluation}
\label{supp:ci_eval}

\paragraph{Prompts.} 
The 100 switch pairs used in Sec.~\ref{sec:analysis} were generated with GPT-5.5~\cite{openai2026gpt55} and cover eight categories of change: human motion, animal motion, object motion, physical state, color, scene, facial expression and lighting.
Each pair consists of a prompt $\yold$ that describes the state before the change and a prompt $\ynew$ that describes the state after it, with the subject and the setting held fixed so that only the changed attribute differs between the two.
Every prompt ends with the suffix ``Realistic video, natural lighting, static camera.'', which is omitted in Tab.~\ref{tab:ci_prompts}.
Note that this set does not overlap with the T2V-CompBench prompts used in Sec.~\ref{sec:exp}.

\begin{table}[h!]
\centering
\caption{\textbf{Prompt pairs of the context inertia set.} Examples of one pair per category.}
\label{tab:ci_prompts}
\footnotesize
\begin{tabular}{l p{0.7\linewidth}}
\toprule
\textbf{Category} & \textbf{Prompt pair example} \\
\midrule
Human motion
& $\yold$: A young man in a gray hoodie is walking slowly in a city park. \newline
  $\ynew$: A young man in a gray hoodie is running fast in a city park. \\
\midrule
Animal motion
& $\yold$: A brown horse is walking slowly in an open meadow. \newline
  $\ynew$: A brown horse is running at full speed in an open meadow. \\
\midrule
Object motion
& $\yold$: A wooden spinning top is spinning upright on a table. \newline
  $\ynew$: A wooden spinning top is lying on its side, not moving on a table. \\
\midrule
Physical state
& $\yold$: A rose is a tightly closed bud in a garden. \newline
  $\ynew$: A rose is fully bloomed with open petals in a garden. \\
\midrule
Color
& $\yold$: A white teddy bear is sitting on a bed. \newline
  $\ynew$: A black teddy bear is sitting on a bed. \\
\midrule
Scene
& $\yold$: A man is standing on a city street on a bright sunny day. \newline
  $\ynew$: A man is standing on a city street in heavy rain. \\
\midrule
Expression
& $\yold$: A close-up of a little boy smiling happily in a classroom. \newline
  $\ynew$: A close-up of a little boy crying with tears on the cheeks in a classroom. \\
\midrule
Lighting
& $\yold$: A white candle is burning with a bright flame on a dark table. \newline
  $\ynew$: A white candle is extinguished with a thin trail of smoke on a dark table. \\
\bottomrule
\end{tabular}
\end{table}

\paragraph{Metrics.}
We use a VLM judge to decide whether a prompt switch takes effect.
After the switch, we slide a 1.5\secs window (two chunks) over the new segment one chunk at a time.
For each window we ask Qwen3-VL-8B~\cite{bai2025qwen3}
``Is this visible after the switch frames: \texttt{<new prompt>}?''
and read the probability of ``Yes'' from the first answer token.
A switch counts as \emph{reached} if any window exceeds $0.5$.

For the context inertia set, we validate the Qwen3-VL-8B judge against 100 human-labeled videos, with annotations blind to the method and the judge's decision.
We confirm that the judge agrees with the human labels on 86\% of the videos (Cohen's $\kappa$~\cite{cohen1960coefficient} = 0.72).

\subsection{Category-wise Results}
\label{supp:ci_category}
\begin{figure}[t]
  \centering
  \includegraphics[width=\linewidth]{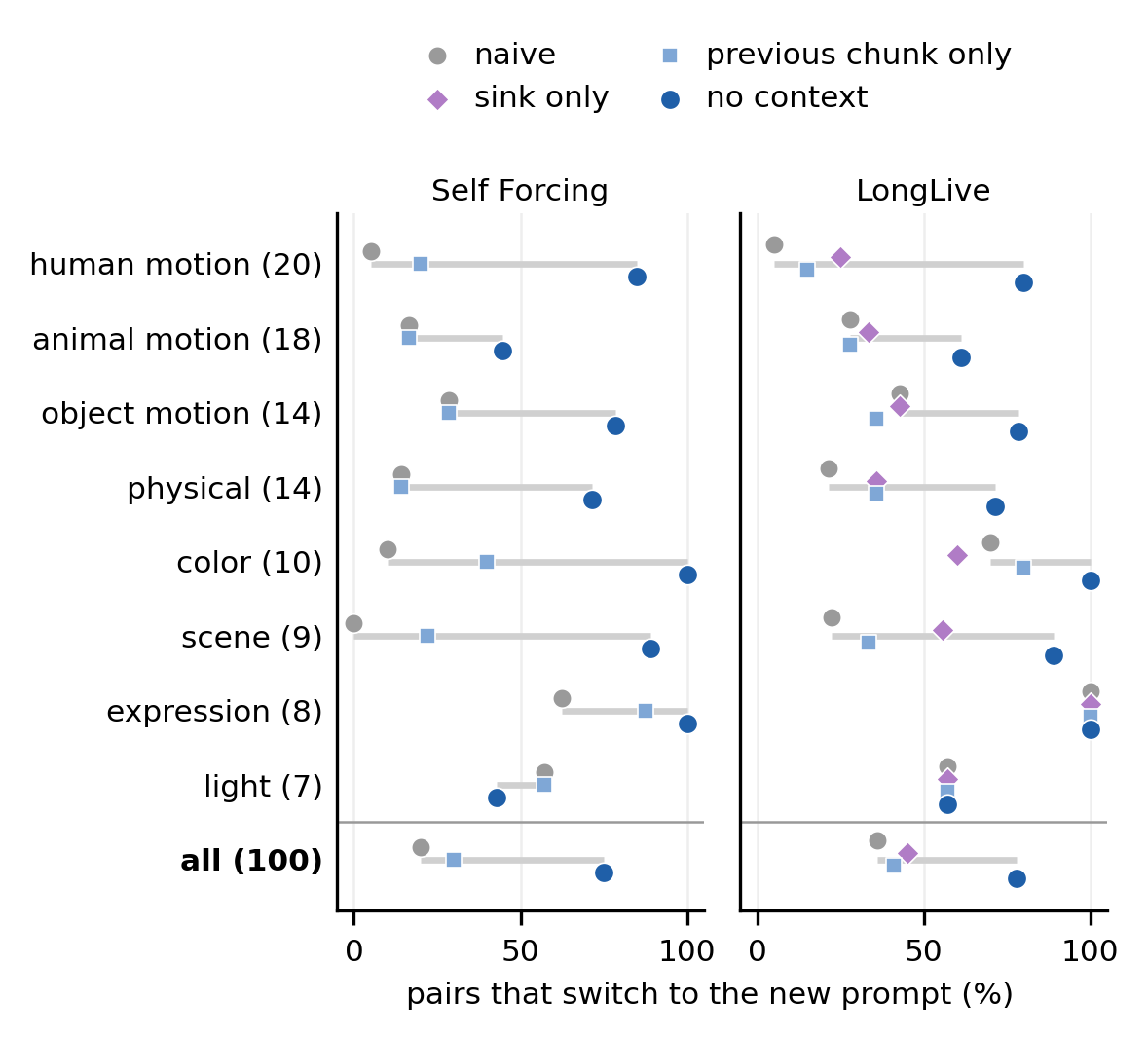}
  \caption{\textbf{Context inertia by category.} Fraction of pairs that switch to the new prompt when the switch chunk is regenerated with part of the cache hidden, on Self Forcing (left) and LongLive (right). \emph{Naive} denotes the base model. The number of pairs per category is in parentheses. Hiding the whole context recovers most switches, whereas a single cached chunk or the sink alone recovers little.}
  \label{fig:category}
\end{figure}

Sec.~\ref{subsec:cause} measures how far the context moves the step-1 prediction. 
Here we ask whether this translates into the generated video, and whether it depends on the kind of change. 
We use the 100 pairs of the context inertia set, which span eight categories of change, and regenerate only the switch chunk with part of the cache hidden at every denoising step: no context, the previous chunk only, or, for LongLive, the sink only. 
Everything else is unchanged, and a pair counts as switched if any two-chunk window after the switch is judged to show the new prompt.

Fig.~\ref{fig:category} shows that without intervention only 20\% of the pairs switch on Self Forcing and 36\% on LongLive.
Hiding the entire context raises these rates to 75\% and 78\%, whereas retaining only the previous chunk (30\% and 41\%) or only the sink (45\% on LongLive) recovers little.
The largest gains occur for human motion, scene and color changes, where the new prompt modifies one property of an otherwise identical scene.
By contrast, expression changes are often realized even with the full context, while lighting changes show no improvement from context removal, suggesting a limitation of the capability of base models rather than context inertia.

\subsection{Step-wise Analysis of Context Inertia}
\label{supp:ci_step}

\begin{figure}[t]
  \centering
  \includegraphics[width=\linewidth]{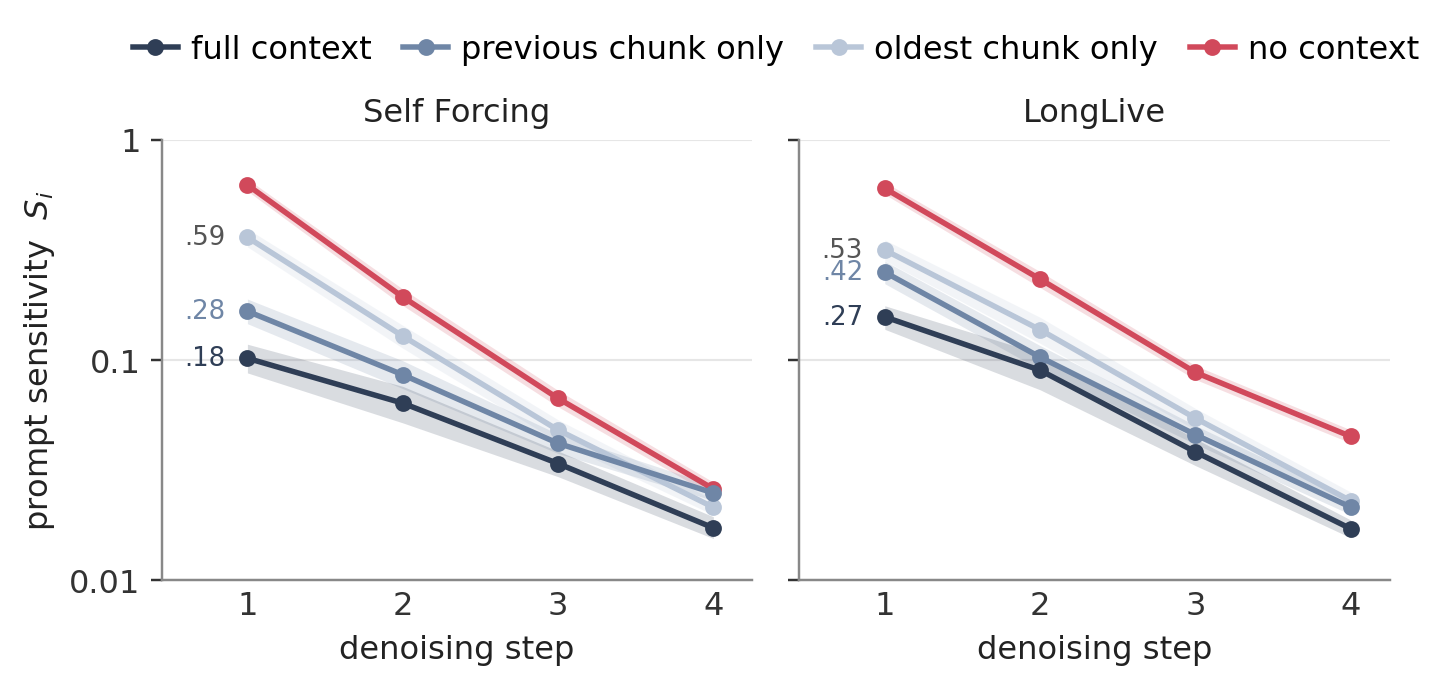}
  \caption{\textbf{Step-wise prompt sensitivity under visible context.} At each denoising step $i$ of the switch chunk, the noisy input is fixed and the prompt sensitivity $S_i(\mathcal{C})$ is measured under each visible context $\mathcal{C}$ on 100 pairs (log scale, shaded bands show 95\% bootstrap confidence intervals). Numbers next to the step-1 points give $\rho_1$, the values of Fig.~\ref{fig:ci_b}. ``Oldest chunk only'' keeps only the oldest chunk still cached, which is chunk 1 for Self Forcing, whose rolling window has dropped chunk 0, and the sink (chunk 0) for LongLive.}
  \label{fig:stepwise}
\end{figure}

In Sec.~\ref{subsec:cause}, we measured the prompt response at the first denoising step. 
Here we repeat the measurement at every step and show that the first step is where the prompt matters most and where the context suppresses it most. 
All results use the 100 pairs of the context inertia set following the same protocol. 
Differences between conditions are tested with a paired bootstrap over pairs (5000 resamples), with 95\% confidence intervals.

\paragraph{Setup.}
At each step $i$ of the switch chunk, the model is run unchanged up to that step and the noisy input $x_k^{t_i}$ is fixed. 
We report the numerator of Eq.~\eqref{eq:rho}, the displacement that the prompt switch causes in the clean prediction, relative to the chunk norm,
\begin{equation}
  S_i(\mathcal{C}) = \frac{\|\hat{x}^{(i)}(\mathcal{C}, \ynew) - \hat{x}^{(i)}(\mathcal{C}, \yold)\|}{\|s_{k-1}\|},
\end{equation} 
so that $\rho_i(\mathcal{C}) = S_i(\mathcal{C}) / S_i(\varnothing)$ for each pair, and Fig.~\ref{fig:ci_b} shows $\rho_1$ averaged over pairs.
We plot $S_i$ rather than $\rho_i$ because the context-free sensitivity itself collapses after step 1, so the ratio approaches 1 at later steps even though the prompt moves the prediction little with or without context. Fig.~\ref{fig:stepwise} reports $S_i$ under four visible contexts.

\paragraph{The prompt acts at the first step, and the context suppresses it there.}
Without context, prompt sensitivity is concentrated at step 1: $S_1(\varnothing)$ is 0.62 on Self Forcing and 0.60 on LongLive, falling by 24$\times$ and 13$\times$ by step 4.
With the full context, step-1 sensitivity drops to 0.10 and 0.16, and the per-pair ratio averages to $\rho_1 = 0.18$ and 0.27.
So, step 1 alone accounts for about 75\% and 67\% of the total sensitivity lost to the context over four steps.
Although relative suppression eases at later steps, the absolute prompt sensitivity has already dropped substantially by step 2 (0.19 and 0.23), leaving little opportunity to recover the new prompt response.

\paragraph{A single cached chunk suffices.}
At step 1, the sensitivity recovers monotonically as the context is reduced. 
With only the previous chunk visible, $\rho_1$ is 0.28 on Self Forcing and 0.42 on LongLive, and with only the first chunk, the farthest from the switch, it is 0.59 and 0.53. 
Every adjacent difference is significant (full to previous only $+0.11$ and $+0.16$, previous only to first only $+0.31$ and $+0.10$, first only to none $+0.41$ and $+0.47$, all $p < 0.001$). 
The immediately preceding chunk is the strongest single source of suppression, yet even one cached chunk halves the response. 
This is why we choose the residual to be restored without any cached context rather than without the most recent chunk only.

\section{Method Details}
\label{supp:method}

\subsection{Rationale for Residual Guidance}
\label{supp:theory}

We provide an approximate interpretation of the residual in Eq.~\eqref{eq:residual} in terms of prompt-conditioned score differences.
This interpretation explains how the residual captures the change induced by the new prompt without conditioning on the cached history, and clarifies the approximations involved.

\paragraph{Posterior factorization.}
Eq.~\eqref{eq:posterior} follows from Bayes' rule applied to
$p(s_k, \ynew \mid x_k^t, \mathcal{C}_{<k})$, together with the conditional independence of the noisy state and the prompt given the clean chunk and cache:
\begin{equation}
p(\ynew \mid s_k, x_k^t, \mathcal{C}_{<k})
=
p(\ynew \mid s_k, \mathcal{C}_{<k}),
\label{eq:condindep}
\end{equation}
which holds because $x_k^t$ is obtained by independently noising $s_k$.
Taking the logarithm separates the posterior into a context term,
$\log p(s_k \mid x_k^t, \mathcal{C}_{<k})$, and a prompt term,
$\log p(\ynew \mid s_k, \mathcal{C}_{<k})$.
The latter remains conditioned on the cache.
Rather than estimating this cache-conditioned term directly, our Residual Guidance constructs a cache-free approximation of the prompt-induced change, motivated by the suppression observed in Sec.~\ref{sec:analysis}.

\paragraph{The residual as a score difference.}
Let $z = \tilde{x}_{k-1}^{\tau}$ and let $p_\tau(z \mid y)$ denote the density of noisy chunks at noise level $\tau$, obtained by noising clean chunks generated under prompt $y$ without cached context.
For an ideal denoiser that predicts the posterior mean, Tweedie's formula~\cite{efron2011tweedie} gives
\begin{equation}
\hat{x}_\theta(z \mid \varnothing, y)
\approx
\frac{
z + \sigma_\tau^2 \nabla_z \log p_\tau(z \mid y)
}{
\alpha_\tau
}.
\label{eq:tweedie}
\end{equation}
The approximation accounts for the fact that the student is not an exact posterior-mean estimator.

At the switch chunk $k$, the preceding chunk $s_{k-1}$ was generated under the old prompt $\yold$.
We approximate its cache-free reconstruction under $\yold$ as
\begin{equation}
\hat{x}_\theta(z \mid \varnothing, \yold)
\approx s_{k-1}.
\label{eq:old-reconstruction}
\end{equation}
Substituting this approximation into Eq.~\eqref{eq:residual} and applying Eq.~\eqref{eq:tweedie} under both prompts yields
\begin{align}
r_k
&\approx
\hat{x}_\theta(z \mid \varnothing, \ynew)
-
\hat{x}_\theta(z \mid \varnothing, \yold)
\nonumber\\
&=
\frac{\sigma_\tau^2}{\alpha_\tau}
\left(
\nabla_z \log p_\tau(z \mid \ynew)
-
\nabla_z \log p_\tau(z \mid \yold)
\right)
\nonumber\\
&=
\frac{\sigma_\tau^2}{\alpha_\tau}
\nabla_z
\log
\frac{p_\tau(z \mid \ynew)}
     {p_\tau(z \mid \yold)}.
\label{eq:score-diff}
\end{align}
Under a common prior over prompts, the last expression is equivalently the gradient of the log posterior odds of $\ynew$ against $\yold$, since the shared marginal $p_\tau(z)$ cancels by Bayes' rule.
Thus, the residual approximates a cache-free, prompt-directed displacement relative to the old prompt, rather than the response to an empty prompt.
It provides a local surrogate for the prompt term in Eq.~\eqref{eq:posterior}, but is not an exact gradient of the cache-conditioned likelihood.

\paragraph{Approximations and scope.}
The interpretation above involves three approximations.
First, Eq.~\eqref{eq:tweedie} assumes that the student prediction approximates the posterior mean under the cache-free noisy marginal.
This is an idealized interpretation of the learned student denoiser rather than an exact identity for a distilled model.
Second, Eq.~\eqref{eq:old-reconstruction} assumes that the cache-free reconstruction under $\yold$ recovers $s_{k-1}$.
However, $s_{k-1}$ was generated with the preceding cache, whereas $p_\tau(\cdot \mid \yold)$ describes the cache-free marginal.
Thus, $s_{k-1}$ need not be a typical sample from this marginal, and the old-prompt reconstruction may not recover it exactly.
In addition, a single reconstruction from a high noise level does not invert the noising, so it departs from $s_{k-1}$ even for a typical sample.
The remainder $\hat{x}_\theta(z \mid \varnothing, \yold) - s_{k-1}$ is the difference between RG and the prompt-difference residual in Supp.~\ref{supp:alternatives}, and that comparison shows that it is not negligible.
Eq.~\eqref{eq:score-diff} therefore describes only the prompt-dependent part of the residual.
The noise level $\tau$ balances preserving the preceding chunk as an anchor against allowing the new prompt to modify its content: too little noise limits the possible displacement, whereas too much noise can discard details needed to preserve the ongoing scene.
We select $\tau = 850$ based on the ablation in Sec.~\ref{sec:ablation}.
Third, transferring the residual from the preceding chunk to the next relies on their local visual correspondence, as motivated by context inertia, rather than an exact equivalence between their predictions.
The score-difference interpretation in Eq.~\eqref{eq:score-diff} applies directly to the switch chunk $k$, whose preceding chunk was generated under $\yold$.
For subsequent chunks within the same prompt interval, $s_{j-1}$ is already generated under $\ynew$.
Their residuals therefore no longer represent an old-versus-new prompt contrast.
Instead, they provide local prompt-directed displacements anchored to the latest generated content, sustaining guidance as the cache incorporates chunks generated under the new prompt.

\subsection{Algorithm}
\label{supp:algorithm}

\begin{algorithm}[t]
\caption{Residual guidance after a prompt switch at chunk $k$}
\label{algorithm}
\begin{algorithmic}[1]
\REQUIRE student $\hat{x}_\theta$, steps $t_1 > \dots > t_Q$, schedule $\lambda$, noise level $\tau$, cache $\mathcal{C}_{<k}$, last chunk $s_{k-1}$, new prompt $\ynew$
\STATE $\epsilon \sim \mathcal{N}(0, I)$ \COMMENT{fixed for this prompt interval}
\STATE $j \gets k$
\WHILE{the active prompt remains $\ynew$}
  \STATE $\tilde{x}_{j-1}^{\tau} \gets \alpha_\tau\, s_{j-1} + \sigma_\tau\, \epsilon$ \COMMENT{re-noise previous chunk}
  \STATE $r_j \gets \hat{x}_\theta(\tilde{x}_{j-1}^{\tau} \mid \varnothing, \ynew) - s_{j-1}$ \COMMENT{cache-free reconstruction}
  \STATE $x_j^{t_1} \sim \mathcal{N}(0, I)$
  \FOR{$i = 1, \dots, Q$}
    \STATE $\hat{x}_j \gets \hat{x}_\theta(x_j^{t_i} \mid \mathcal{C}_{<j}, \ynew) + \lambda_i\, r_j$ \COMMENT{guided prediction}
    \IF{$i < Q$}
      \STATE $\epsilon' \sim \mathcal{N}(0, I)$
      \STATE $x_j^{t_{i+1}} \gets \alpha_{t_{i+1}} \hat{x}_j + \sigma_{t_{i+1}} \epsilon'$
    \ENDIF
  \ENDFOR
  \STATE $s_j \gets \hat{x}_j$ and append its keys and values to $\mathcal{C}$
  \STATE $j \gets j + 1$
\ENDWHILE
\end{algorithmic}
\end{algorithm}

\noindent
Alg.~\ref{algorithm} gives the procedure following a prompt switch at chunk $k$, which continues until the next prompt switch.
A single noise sample $\epsilon$ is drawn at the switch and shared across all residuals during the prompt interval (line 1).
For each chunk $j \geq k$, the preceding chunk $s_{j-1}$ is re-noised to level $\tau$ and reconstructed in one cache-free forward pass under the active prompt.
The residual $r_j$ is the displacement of this reconstruction from $s_{j-1}$ (lines 4--5).
The chunk is then denoised using the original sampler, with $\lambda_i r_j$ added to the clean prediction at each step (line 8).
With $\lambda = (1, 0.5, 0, 0)$, the residual is injected only at the first two of the four denoising steps.
The resulting clean chunk $s_j$ is appended to the cache as usual (line 14).
Thus, the residual itself is never written to the cache; only the resulting generated chunk contributes to subsequent context.
When a new prompt arrives, the procedure restarts with a fresh noise sample and the updated prompt.
The procedure requires one additional cache-free forward pass per generated chunk while the prompt remains active, with no additional model evaluations at the other denoising steps.

\section{Evaluation Protocol}
\label{supp:eval}

\subsection{Benchmarks and Prompt Construction}
\label{supp:benchmarks}
\paragraph{Prompt Details.}
The prompts of T2V-CompBench dynamic attribute~\cite{sun2025t2v} describe a change of attribute in a single sentence, for example ``A timelapse of a leaf transitioning from green to bright red as autumn progresses''.
Therefore, we split each prompt into one for the initial state and one for the final state, with the subject and the setting kept fixed, for example $\yold$ ``A green leaf on a tree branch'' and $\ynew$ ``A bright red leaf on a tree branch''.
A shared description of motion, lighting and camera is then appended to both prompts (e.g., ``\dots, swaying gently in the breeze, soft natural light, slow camera push-in'').
This description serves two purposes.
First, a bare description of a state tends to yield a nearly static video, whereas the appended description gives the scene ongoing motion that must continue across the switch, as in interactive use.
Second, the streaming models are trained on long, detailed prompts, and the appended description keeps the inputs close to this distribution.
Since the description is identical in $\yold$ and $\ynew$, the two prompts differ only in the changed state, so any difference after the switch can be attributed to it.
The split was performed with an LLM and checked manually by the authors.
StoryEval~\cite{wang2024storyeval} contains prompts with two to four events.
Since a two-event prompt is a single switch, which the dynamic attribute setting already covers, we randomly sample 100 three-event prompts.
InteractiveBench~\cite{yang2025longlive} is used as released by MemFlow~\cite{ji2025memflow}, with each of the six prompts active for 10\secs.

\subsection{Metric Details}
\label{supp:metrics}

All three settings share the same metrics. For each switch we call the chunks generated under the new prompt the post-switch segment, and quality metrics that measure consistency across frames exclude the switch chunk from it, so that the change requested by the prompt is not penalized as an inconsistency. 

\paragraph{Achievement.}
A VLM judge (Qwen3-VL-8B~\cite{bai2025qwen3}) compares the last pre-switch frame with a sliding window of two chunks (8 frames, $832{\times}480$) over the post-switch segment, and is asked whether the requested change
(e.g., ``a green leaf'' $\rightarrow$ ``a red leaf'') is visible. A switch is reached if the response gets $P(\text{yes}) > 0.5$.
Since a model may be unable to depict the new state at all, we generate an upper-bound video from the same seed with the new prompt as the only prompt, and Achievement is the fraction of reached switches among those for which this upper bound is reached.
If a video faces several switches, Achievement is averaged over them. 
We discuss the reliability of this metric more in Supp.~\ref{supp:judge_reliability}.

\paragraph{T2V-CompBench dynamic attribute.}
On the dynamic attribute setting we additionally run the official evaluator of T2V-CompBench~\cite{sun2025t2v}, which uses LLaVA-1.6-34B~\cite{liu2023visual} to score per frame whether the attribute changes from the initial to the final state, on the whole video and all 200 pairs. 
The score is close to 0 for a video that never changes and small in absolute value by construction.

\paragraph{CLIP.}
CLIP similarity~\cite{radford2021learning} is measured between the frames of the post-switch segment and the new prompt, averaged over chunks and reported $\times 100$. 
On InteractiveBench it is computed per 10\secs segment against the prompt of that segment, following the report form of previous works.

\paragraph{Quality.}
We follow the VBench~\cite{huang2023vbench} quality score exactly. The seven dimensions are each min-max normalized to $[0, 1]$ with the official VBench constants, and the quality score is their weighted mean with weight 0.5 on dynamic degree and 1 on the other six, divided by the total weight of 6.5. Subject consistency (DINO ViT-B/16~\cite{caron2021emerging} feature similarity across frames), background consistency (CLIP feature similarity across frames), temporal flickering (mean absolute pixel difference between frames), motion smoothness (difference between AMT~\cite{li2023amt} frame interpolation and the generated frames), aesthetic quality (LAION~\cite{schuhmann2022laion} aesthetic predictor on CLIP ViT-L/14) and dynamic degree (RAFT~\cite{teed2020raft} optical flow) are computed on the post-switch segment excluding the switch chunk, and imaging quality (MUSIQ~\cite{ke2021musiq}) on the whole post-switch segment. Background consistency is computed only on pairs in which the prompt does not change the background (150 of 200 on dynamic attribute). On InteractiveBench the quality score is additionally computed with VBench-Long on the whole video (Supp.~\ref{supp:vbench_long}).

\paragraph{Overall.}
$(4 \times \text{Quality} + \text{Achievement}) / 5$, which is the VBench total score with Achievement in place of the semantic score.

\subsection{Reliability of the VLM Judge}
\label{supp:judge_reliability}

Achievement in all tables is determined by Qwen3-VL-8B~\cite{bai2025qwen3}.
To assess whether our results are sensitive to this choice of judge, we repeat the evaluation with Qwen3-VL-32B, a larger variant of the same model, and InternVL3.5-38B~\cite{wang2025internvl3}.
We randomly sample 100 prompt pairs from the dynamic attribute setting, distributed across the seven base models, and evaluate both the baseline and residual-guided videos for each pair, yielding 200 videos in total.
The three judges achieve pairwise agreement rates of 91 to 93\%, with Cohen's $\kappa$~\cite{cohen1960coefficient} ranging from 0.81 to 0.86 across judge pairs.

We further compare the three judges against human annotations on the same 200 videos.
To avoid bias, the videos are presented in random order without revealing the base model or whether residual guidance was used.
The annotator views each video in full, whereas the VLM judges evaluate the post-switch segment using a sliding window of two chunks.
The 8B, 32B, and InternVL judges achieve 84\%, 85\%, and 85\% agreement with the human labels, respectively, with Cohen's $\kappa$ values of 0.67, 0.68, and 0.70.
These results show that all three judges are similarly aligned with human judgments, confirming that the 8B judge used in the main tables is not an outlier.

It is also important to note that the improvement in prompt fidelity is supported by metrics that do not rely on the same VLM judge.
CLIP, which does not involve a VLM, and the official dynamic attribute score of T2V-CompBench, which uses a different model (LLaVA-1.6-34B~\cite{liu2023visual}) and evaluation protocol, both improve for every base model in Tab.~\ref{tab:dynattr}, consistent with Achievement.

\subsection{User Study Details}
\label{supp:user_study}

\paragraph{Pairs.}
The study uses the dynamic attribute setting of Tab.~\ref{tab:dynattr} with its seven base models.
Each item compares a base model and the same model with residual guidance on the same prompt pair and seed, so that the two videos are identical before the switch.
We randomly select 15 pairs per base model (105 in total).

\paragraph{Presentation.}
Each item is a single image with two rows, one per video, labeled Top and Bottom.
A row shows five frames at $832\times480$: the middle frame of the last chunk before the switch (frame 26), a red bar marking the switch (frame 33, 2.06\,s at 16\,fps), and the middle frames of the four chunks after it (frames 38, 50, 62 and 74).
Above the frames we print only the part of the prompt that changes, as Before and After.
The method behind each row is hidden.
Residual guidance is on top in half of the items of each form at random positions, and the item order is shuffled per form.

\paragraph{Questions.}
For each item, raters answer three questions.
Q1 asks, for each video separately, whether they noticed any unwanted change after the marker apart from the change the prompt asks for, such as in the background or other objects (No or Yes).
Q2 asks which video better shows the new prompt after the marker (Top, Bottom or No difference).
Q3 asks which video they would rather get if they were changing the prompt live while it is generating (Top, Bottom or No difference).
The instructions state that a change requested by the prompt does not count as an unwanted change.

\paragraph{Forms and participants.}
The user study does not involve any sensitive personal information or foreseeable risks to participants.
All participants provided informed consent before participation.
The 105 pairs are split into five forms of 21 pairs, 3 per base model, and each participant completes one form.
We recruited 20 participants who were not involved in the project, three to five per form.
Each form took 10 to 15 minutes.

\paragraph{Analysis.}
For Q2 and Q3, a judgment counts 1 when residual guidance is chosen, 0 when the base model is chosen and 0.5 for no difference.
For Q1, it counts 1 when only the base model's video is flagged, 0 when only ours is flagged and 0.5 otherwise.
We report the mean over the 420 judgments, and 95\% bootstrap confidence intervals over raters and pairs (2000 resamples) are 73.1\% [66.8, 79.2] for Q2, 70.7\% [64.7, 76.9] for Q3 and 44.3\% [38.9, 49.6] for Q1.

\section{Additional Results}
\label{supp:more_res}

\subsection{Effect of the Requested Change on Quality}
\label{supp:quality}
\begin{table*}[h!]
\centering
\caption{\textbf{VBench dimensions on T2V-CompBench dynamic attribute.} Each cell gives the base model $\rightarrow$ the same model with residual guidance, computed on the post-switch segment as in Tab.~\ref{tab:dynattr}.}
\label{tab:vbench_dims}
\small
\setlength{\tabcolsep}{5pt}
\resizebox{\textwidth}{!}{%
\begin{tabular}{l cccccccc}
\toprule
\textbf{Method} & \textbf{Subject} & \textbf{Background} & \textbf{Flickering} & \textbf{Smoothness} & \textbf{Aesthetic} & \textbf{Imaging} & \textbf{Dynamic} & \textbf{Quality} \\
\midrule
Self Forcing   & 0.987 $\rightarrow$ 0.960 & 0.985 $\rightarrow$ 0.967 & 0.986 $\rightarrow$ 0.977 & 0.992 $\rightarrow$ 0.988 & 0.578 $\rightarrow$ 0.566 & 67.09 $\rightarrow$ 66.80 & 0.250 $\rightarrow$ 0.380 & 0.813 $\rightarrow$ 0.806 \\
Causal Forcing & 0.981 $\rightarrow$ 0.956 & 0.982 $\rightarrow$ 0.965 & 0.977 $\rightarrow$ 0.962 & 0.988 $\rightarrow$ 0.978 & 0.584 $\rightarrow$ 0.572 & 68.22 $\rightarrow$ 67.92 & 0.425 $\rightarrow$ 0.600 & 0.821 $\rightarrow$ 0.813 \\
MemFlow        & 0.987 $\rightarrow$ 0.976 & 0.984 $\rightarrow$ 0.976 & 0.985 $\rightarrow$ 0.980 & 0.992 $\rightarrow$ 0.989 & 0.571 $\rightarrow$ 0.565 & 67.51 $\rightarrow$ 67.63 & 0.175 $\rightarrow$ 0.235 & 0.805 $\rightarrow$ 0.802 \\
\midrule
LongLive       & 0.987 $\rightarrow$ 0.978 & 0.986 $\rightarrow$ 0.978 & 0.988 $\rightarrow$ 0.982 & 0.993 $\rightarrow$ 0.989 & 0.580 $\rightarrow$ 0.571 & 66.37 $\rightarrow$ 66.57 & 0.145 $\rightarrow$ 0.215 & 0.805 $\rightarrow$ 0.802 \\
Infinity-RoPE  & 0.990 $\rightarrow$ 0.968 & 0.985 $\rightarrow$ 0.971 & 0.991 $\rightarrow$ 0.983 & 0.993 $\rightarrow$ 0.990 & 0.579 $\rightarrow$ 0.571 & 65.14 $\rightarrow$ 64.06 & 0.155 $\rightarrow$ 0.310 & 0.806 $\rightarrow$ 0.803 \\
\midrule
Anchor Forcing & 0.978 $\rightarrow$ 0.965 & 0.979 $\rightarrow$ 0.969 & 0.975 $\rightarrow$ 0.965 & 0.987 $\rightarrow$ 0.980 & 0.578 $\rightarrow$ 0.578 & 67.12 $\rightarrow$ 67.06 & 0.365 $\rightarrow$ 0.540 & 0.811 $\rightarrow$ 0.813 \\
SWIFT          & 0.990 $\rightarrow$ 0.978 & 0.988 $\rightarrow$ 0.979 & 0.990 $\rightarrow$ 0.984 & 0.994 $\rightarrow$ 0.990 & 0.580 $\rightarrow$ 0.573 & 66.47 $\rightarrow$ 66.71 & 0.120 $\rightarrow$ 0.150 & 0.806 $\rightarrow$ 0.799 \\
\bottomrule
\end{tabular}%
}
\end{table*}
\begin{figure*}[h!]
  \centering
  \includegraphics[width=\textwidth]{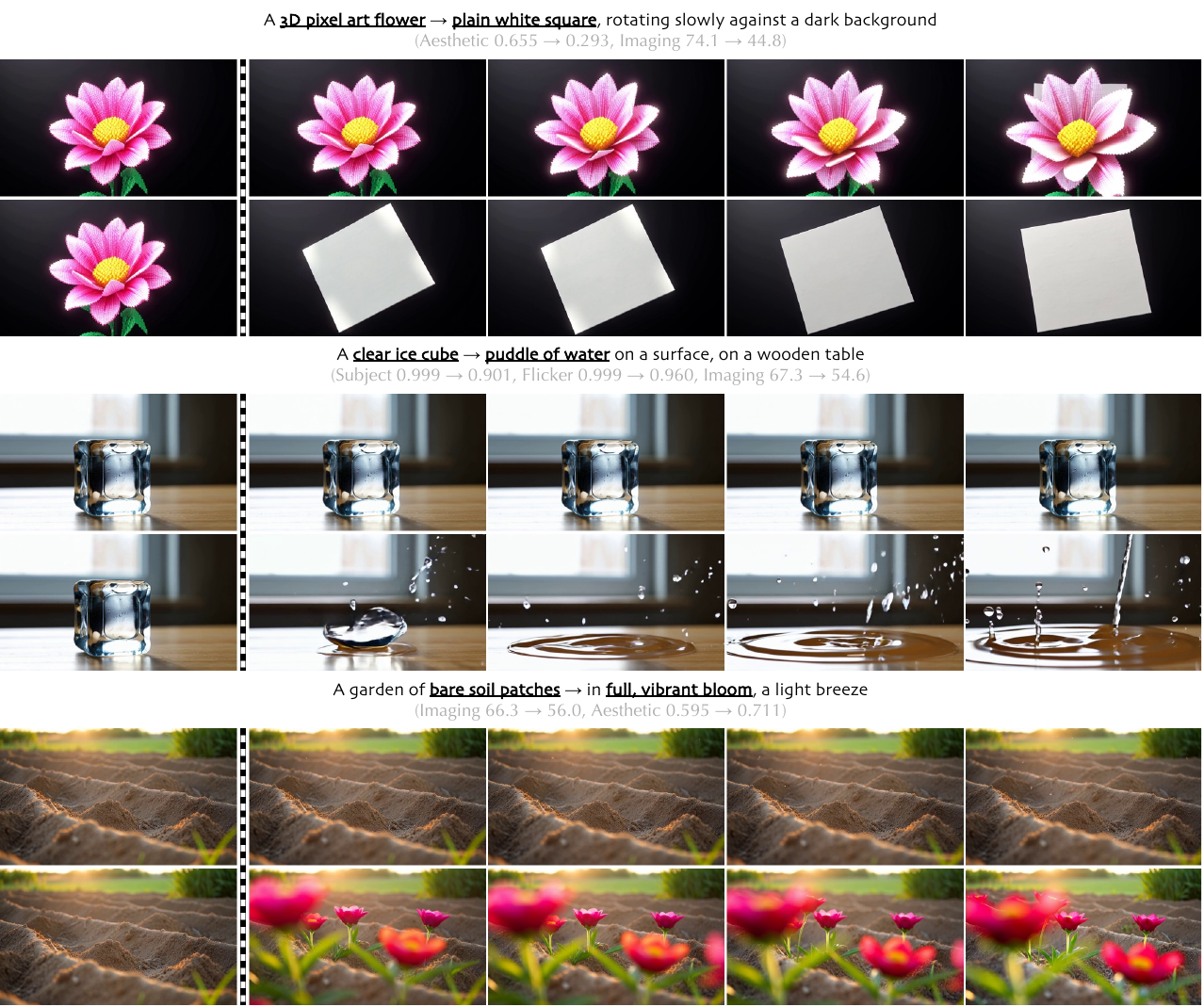}
  \caption{\textbf{VBench Quality penalizes the requested change.}
  In each example, the top row is the base model and the bottom row is \ours, and the dashed bar marks the prompt switch.
  The examples are from LongLive (top), Causal Forcing (middle) and Infinity-RoPE (bottom).
  The gray numbers give the VBench dimension scores of the base model and of \ours.
  \ours\ follows the new prompt while the base model keeps the old state, yet \ours\ scores lower on aesthetic quality, subject consistency and imaging quality, respectively.}
  \label{fig:quality}
\end{figure*}
\ours\ changes VBench Quality by at most 0.008, with a small decrease on six of the seven models (Tab.~\ref{tab:dynattr}).
Tab.~\ref{tab:vbench_dims} breaks this score into its seven dimensions.
\ours\ lowers the three consistency dimensions (subject consistency, background consistency and temporal flickering) by up to 0.03 on every model and raises dynamic degree, whereas aesthetic quality, imaging quality and motion smoothness stay within about 0.01 of the base model.
The two movements have the same cause.
The base model keeps the initial state after the switch and is therefore nearly static, which yields high consistency for free~\cite{huang2023vbench}, while \ours\ realizes the change and the frames across it differ.
The scene outside the changed object remains substantially closer to the pre-switch frame than in the no-context setting (Unint.\ in Tab.~\ref{tab:alternatives}), although some scene drift remains (Supp.~\ref{supp:switch_chunk}).
Thus, the drop in consistency is partly associated with the requested change, but does not rule out unintended changes to the surrounding scene.

To examine individual cases, we inspect the prompt pairs on which Quality drops the most.
Fig.~\ref{fig:quality} shows three of them, together with the VBench dimensions that change the most in either direction.
The ice cube example (middle) mirrors the average trend.
When the ice cube melts into a puddle of water, subject consistency falls from 0.999 to 0.901 and temporal flickering from 0.999 to 0.960.
The other two examples involve the aesthetic and imaging scores, which are unchanged on average but depend on the content that is shown.
When a 3D pixel art flower turns into a plain white square as requested, aesthetic quality falls from 0.655 to 0.293 and imaging quality from 74.1 to 44.8.
When bare soil turns into a garden in full bloom, imaging quality falls from 66.3 to 56.0 while aesthetic quality rises from 0.595 to 0.711.
These scores thus follow the requested content in either direction, consistent with their negligible change on average.
In all three cases, the base model keeps its high scores only because the video does not change.
The decrease in Quality is therefore largely a consequence of the requested change, and Quality should be read together with Achievement, which is the reason Overall combines the two.

\subsection{Scene Preservation at the Switch}
\label{supp:switch_chunk}

\begin{figure*}[h!]
  \centering
  \includegraphics[width=\textwidth]{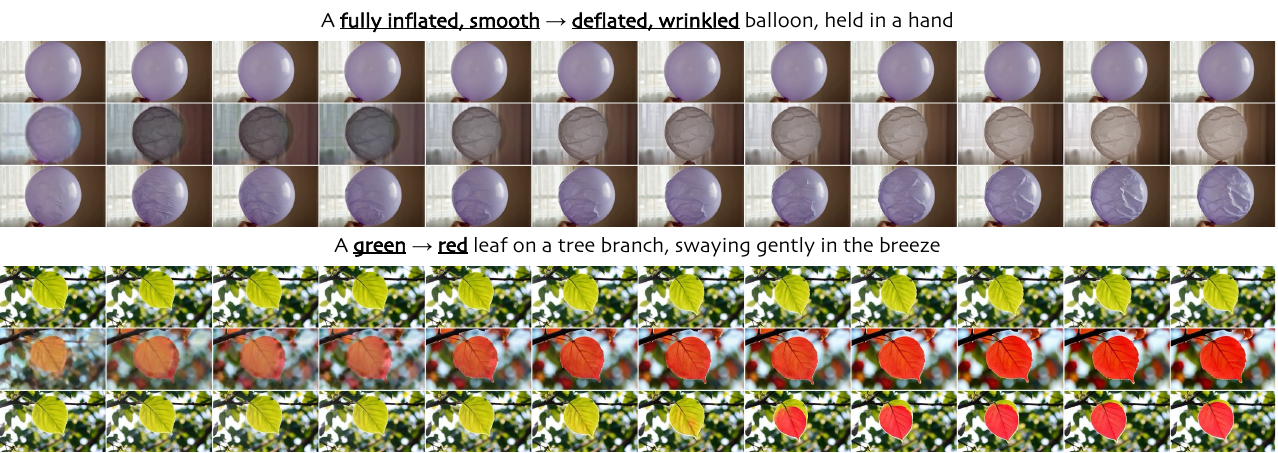}
  \caption{\textbf{Frames of the switch chunk.} LongLive (top) and Anchor Forcing (bottom). In each example, the first row is the base model, which continues the previous state. The second row empties the cache at the switch, as in the no-context condition of Sec.~\ref{sec:analysis}, and follows the new prompt but loses the scene in the switch chunk. The third row is \ours, which realizes the new state within the switch chunk while preserving the subject and background.}
  \label{fig:switch}
\end{figure*}

\begin{table}[t]
\centering
\caption{\textbf{Scene preservation on T2V-CompBench dynamic attribute.}
Distance between the last pre-switch frame and the post-switch frames over all 200 prompt pairs, at the switch chunk (\emph{Switch}) and averaged over the four post-switch chunks (\emph{Mean}).
Lower is closer to the pre-switch scene.
Structure is the DINO self-similarity distance~\cite{tumanyan2022splicing} on the whole frame.
DreamSim~\cite{fu2023dreamsim} is computed with the changed object masked out (OWLv2~\cite{minderer2023scaling} and SAM~\cite{ke2023segment}).
The reference rows report the same distances between a pre-switch frame and a transformed copy of it, on the whole frame.
Reference values do not depend on the chunk and span both columns.}
\label{tab:scene}
\small
\resizebox{0.8\columnwidth}{!}{%
\begin{tabular}{l cc cc}
\toprule
& \multicolumn{2}{c}{\textbf{Structure $\downarrow$}} & \multicolumn{2}{c}{\textbf{DreamSim (bg) $\downarrow$}} \\
\cmidrule(lr){2-3}\cmidrule(lr){4-5}
\textbf{Method} & Switch & Mean & Switch & Mean \\
\midrule

Self Forcing
& 0.010 & 0.019 & 0.015 & 0.028 \\
\oursrow
+ \ours
& 0.030 & 0.045 & 0.060 & 0.091 \\

Causal Forcing
& 0.023 & 0.037 & 0.028 & 0.045 \\
\oursrow
+ \ours
& 0.043 & 0.059 & 0.081 & 0.117 \\

MemFlow
& 0.009 & 0.020 & 0.011 & 0.024 \\
\oursrow
+ \ours
& 0.017 & 0.033 & 0.021 & 0.041 \\

\midrule
LongLive
& 0.007 & 0.016 & 0.009 & 0.023 \\
\oursrow
+ \ours
& 0.022 & 0.035 & 0.030 & 0.047 \\

Infinity-RoPE
& 0.008 & 0.014 & 0.012 & 0.024 \\
\oursrow
+ \ours
& 0.027 & 0.040 & 0.047 & 0.072 \\

\midrule
Anchor Forcing
& 0.013 & 0.030 & 0.016 & 0.039 \\
\oursrow
+ \ours
& 0.026 & 0.046 & 0.032 & 0.060 \\

SWIFT
& 0.007 & 0.013 & 0.009 & 0.018 \\
\oursrow
+ \ours
& 0.023 & 0.033 & 0.031 & 0.046 \\

\midrule
\textit{Reference} & & & & \\
\quad zoom $1.2\times$ & \spanval{0.047} & \spanval{0.037} \\
\quad shift 20\%       & \spanval{0.110} & \spanval{0.058} \\
\quad another video    & \spanval{0.133} & \spanval{0.782} \\
\bottomrule
\end{tabular}%
}
\end{table}

A stronger switch could simply come from redrawing the scene.
We examine this first inside the switch chunk and then over the post-switch segment.

Fig.~\ref{fig:switch} shows the frames of the switch chunk.
Emptying the cache at the switch, as in the no-context condition of Sec.~\ref{sec:analysis}, follows the new prompt but produces a chunk that is detached from the preceding frames, with the balloon and the leaf redrawn in a different scene.
RG changes the state within the same chunk while retaining more of the preceding subject and background than cache removal, making the switch more immediate without fully disrupting visual continuity.

To quantify this, we measure how far the post-switch frames move from the last pre-switch frame, using a structure distance~\cite{tumanyan2022splicing} (DINO self-similarity) and a perceptual distance (DreamSim~\cite{fu2023dreamsim}) computed with the changed object masked out (Tab.~\ref{tab:scene}).
The mask is obtained by detecting the changing object with OWLv2~\cite{minderer2023scaling}, using the two attribute states of the prompt as text queries, segmenting the detected box with SAM~\cite{ke2023segment}.
\ours\ moves the scene further than the base model on every model.
This is expected, since the base model often does not switch at all and the structure distance also includes the requested change.
The change nevertheless remains bounded.
The mean structure distance with \ours\ is between 0.033 and 0.059, at or below a $1.2\times$ zoom of the same frame (0.047) on six of the seven models and far below that of another video (0.133).
The background distance is between a $1.2\times$ zoom and a 20\% shift (0.037 and 0.058) on LongLive, MemFlow and SWIFT, and up to about twice the shift on the other models, an order of magnitude below another video (0.782).
The switch itself introduces no abrupt break, as the structure distance at the switch chunk stays below the zoom reference on every model.
In contrast, emptying the cache changes the background substantially more than \ours, with an object-masked DreamSim distance of 0.191 and 0.163 on Self Forcing and LongLive against 0.089 and 0.046 for \ours\ (Tab.~\ref{tab:alternatives}).

\subsection{More Qualitative Results}
\label{supp:qualitative}

\begin{figure*}[h!]
\centering
\includegraphics[width=\textwidth]{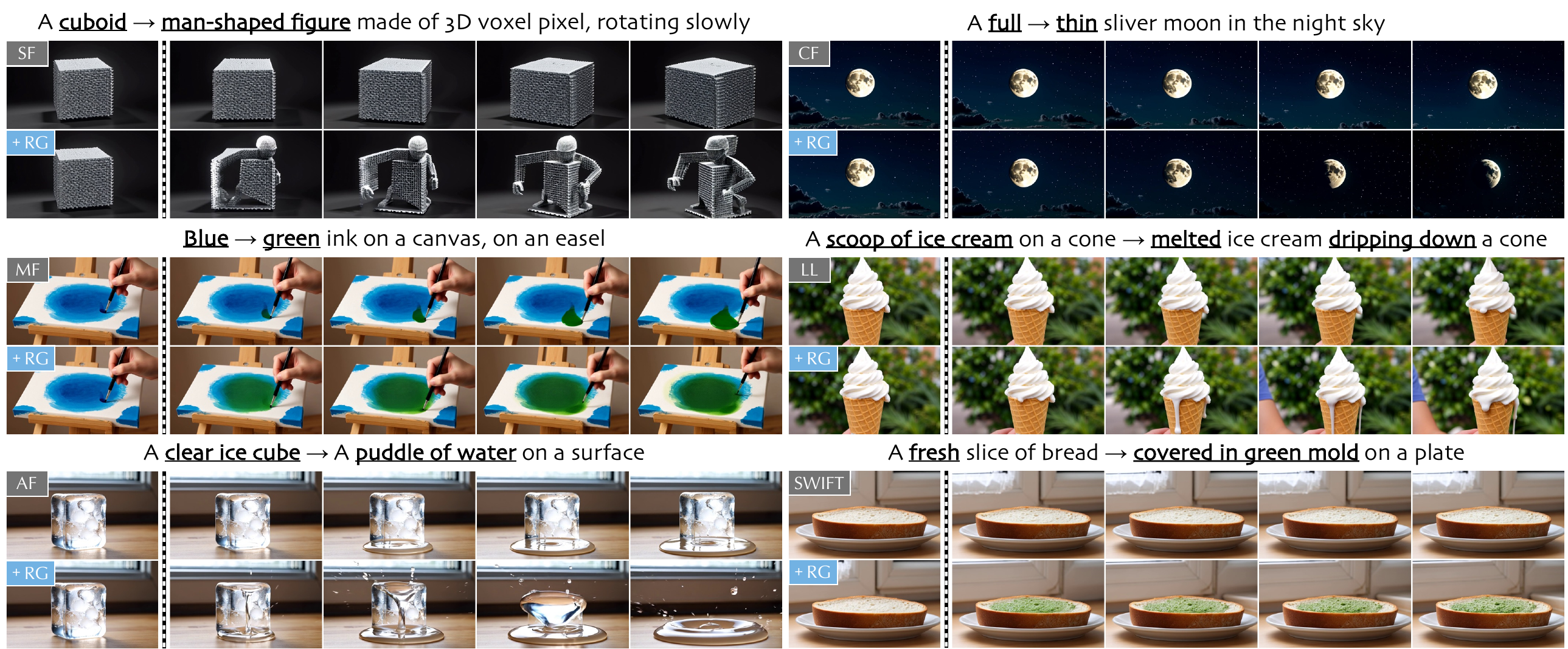}
\caption{\textbf{Qualitative results on T2V-CompBench dynamic attribute.} Each example shows a base model (top) and the same model with RG (bottom). The white-to-black bar marks the prompt switch.}
\label{fig:supp_dynattr}
\end{figure*}

\begin{figure*}[h!]
\centering
\includegraphics[width=\textwidth]{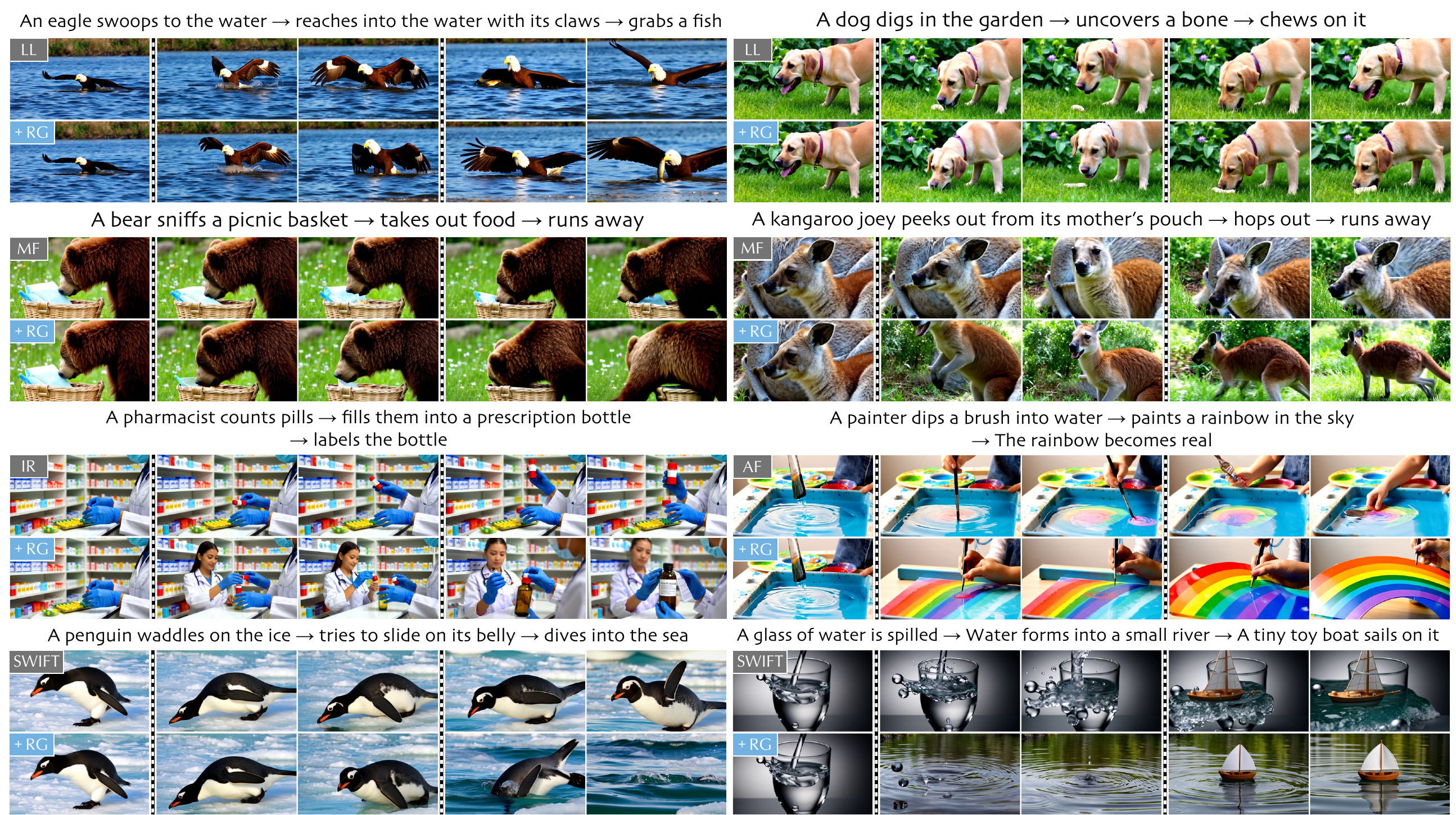}
\caption{\textbf{Qualitative results on StoryEval.} Each example shows a base model (top) and the same model with RG (bottom) over three events. The white-to-black bars mark the two prompt switches.}
\label{fig:supp_storyeval}
\end{figure*}

\begin{figure*}[h!]
\centering
\includegraphics[width=\textwidth]{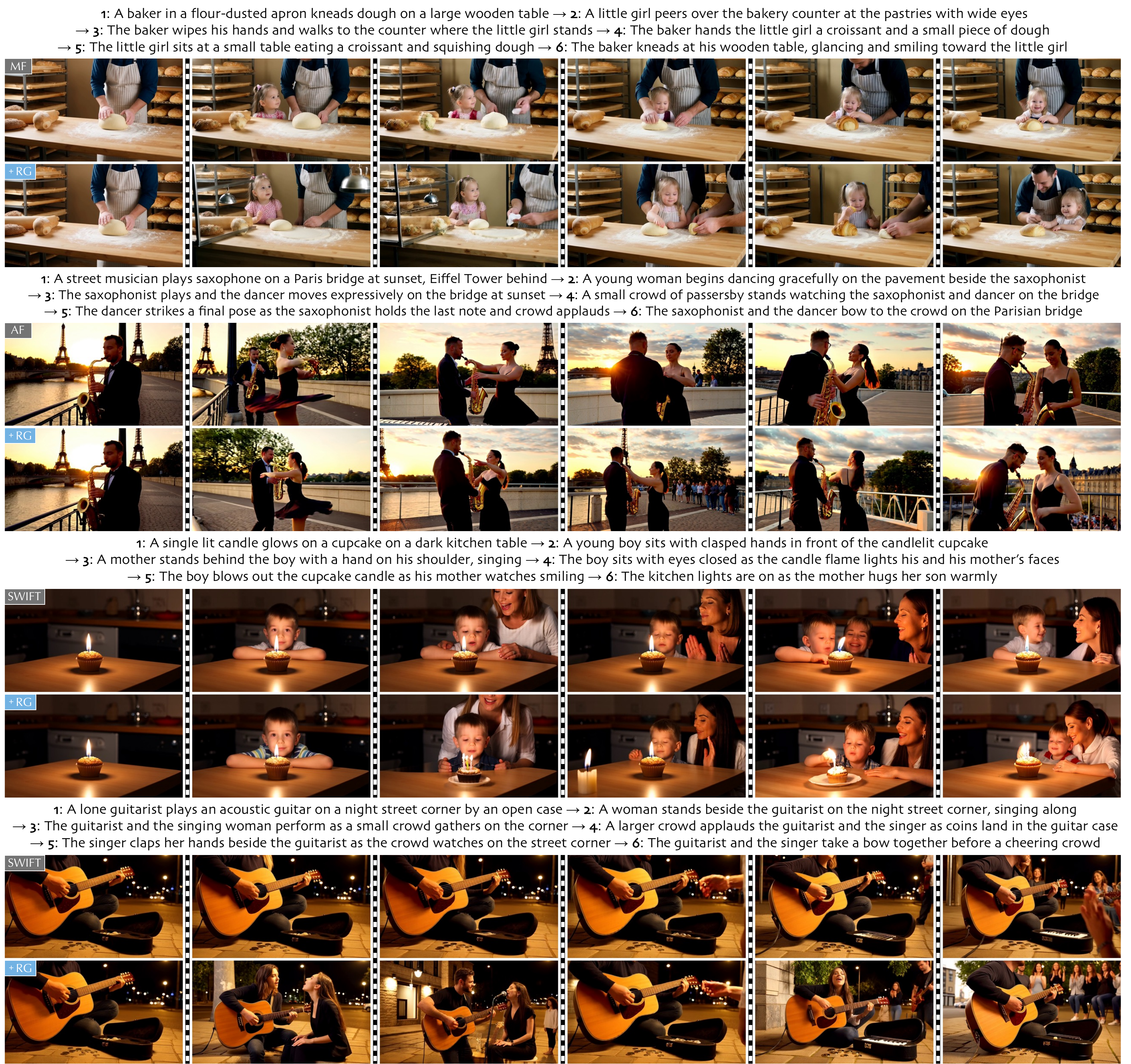}
\caption{\textbf{Qualitative results on InteractiveBench.} Each example shows a base model (top) and the same model with RG (bottom) over six prompts in 60\secs, one frame per prompt. The white-to-black bars mark the five prompt switches.}
\label{fig:supp_interactive}
\end{figure*}

Figs.~\ref{fig:supp_dynattr}, \ref{fig:supp_storyeval} and \ref{fig:supp_interactive} give further examples on the three settings, one base model per example and the same model with RG below it.

On T2V-CompBench dynamic attribute (Fig.~\ref{fig:supp_dynattr}), the base models keep the initial state after the switch, whereas RG realizes changes of shape (the cuboid unfolds into a figure, the ice cube melts into a puddle), state (the ice cream drips, the bread grows mold) and color (the ink turns green) within the first chunks, with the object, its position and the background preserved. 
The thin moon on Causal Forcing shows the change arriving gradually over several chunks rather than at once.

On StoryEval (Fig.~\ref{fig:supp_storyeval}), the base models tend to continue the first event with small variations, as with the bear that never leaves the basket or the pharmacist who keeps counting pills. 
RG follows each event as it is prompted, so the joey hops out and runs, the pharmacist fills and labels the bottle, and the penguin slides and dives. 
Also for SWIFT, the spilled water from the glass forms into a river with RG where the glass remains for the base model.

On InteractiveBench (Fig.~\ref{fig:supp_interactive}), the base models reach some of the six prompts but miss others, for example SWIFT never shows the singer beside the guitarist and the crowd appears only at the end. 
With RG each new subject enters when prompted and the scene develops in step with the script, including the new subject interacting with the existing one (the baker handing the croissant, the mother hugging the boy) while the setting stays continuous across all five switches.

\subsection{VBench-Long on InteractiveBench}
\label{supp:vbench_long}

\begin{table}[t]
\centering
\caption{\textbf{VBench-Long dimensions on InteractiveBench.} Computed on the full 60\secs video (100 videos). Quality is the VBench score over all seven dimensions. Temporal flickering, motion smoothness and aesthetic quality change by less than 0.005 on every model and are omitted.}
\label{tab:vbench_long}
\small
\setlength{\tabcolsep}{8pt}
\resizebox{\linewidth}{!}{%
\begin{tabular}{l cccc c}
\toprule
\textbf{Method} & \textbf{Subject $\uparrow$} & \textbf{Background $\uparrow$} & \textbf{Imaging $\uparrow$} & \textbf{Dynamic $\uparrow$} & \textbf{Quality $\uparrow$} \\
\midrule
MemFlow
& 0.976 & 0.960 & 0.718 & 0.269 & 0.815 \\
\oursrow + \ours
& 0.971 & 0.954 & 0.724 & 0.340 & 0.817 \\
\midrule
LongLive
& 0.975 & 0.960 & 0.719 & 0.280 & 0.818 \\
\oursrow + \ours
& 0.968 & 0.953 & 0.722 & 0.354 & 0.819 \\
Infinity-RoPE~\cite{yesiltepe2026infinity}
& 0.981 & 0.965 & 0.713 & 0.370 & 0.825 \\
\oursrow + \ours
& 0.972 & 0.959 & 0.712 & 0.505 & 0.829 \\
\midrule
Anchor Forcing
& 0.959 & 0.954 & 0.724 & 0.729 & 0.835 \\
\oursrow + \ours
& 0.954 & 0.949 & 0.722 & 0.789 & 0.834 \\
SWIFT
& 0.974 & 0.960 & 0.727 & 0.299 & 0.820 \\
\oursrow + \ours
& 0.966 & 0.954 & 0.725 & 0.378 & 0.820 \\
\bottomrule
\end{tabular}%
}
\end{table}

Tab.~\ref{tab:vbench_long} reports the VBench-Long dimensions on InteractiveBench, computed on the whole 60\secs video.
The values therefore differ from the Quality in Tab.~\ref{tab:longer}, which is computed on the post-switch segments (Supp.~\ref{supp:metrics}).
The pattern is the same across all five base models and mirrors the 5\secs results.
Subject and background consistency decrease by 0.005 to 0.009 with residual guidance, which is expected since each of the five prompt switches is now realized rather than ignored, and a video that changes its content cannot score as high on frame-level consistency as one that does not. 
Imaging quality is unchanged within 0.006 in either direction, and temporal flickering, motion smoothness and aesthetic quality change by less than 0.005 on every model, so the guidance introduces no visible degradation. 
Dynamic degree rises substantially, from 0.27 to 0.34 on MemFlow and from 0.37 to 0.51 on Infinity-RoPE for example, which reflects that the motion follows the changing prompts instead of settling into a static scene. 
Under the official weighting, the aggregated quality score is unchanged or slightly higher with residual guidance on every base model except Anchor Forcing, where it is lower by 0.001.
Under both protocols, Quality stays within 0.006 of the base model.

\subsection{Real-Time Practicality}
\label{supp:speed}
\begin{table}[t]
\centering
\caption{\textbf{Throughput and responsiveness.} Measured at batch size 1 with 5\secs videos and one switch. RG adds one forward pass per chunk, reducing throughput by around 6\%.}
\vspace{-5pt}
\label{tab:practical}
\footnotesize
\setlength{\tabcolsep}{5pt}
\renewcommand{\arraystretch}{0.7}
\setlength{\extrarowheight}{2.5pt}
\resizebox{\linewidth}{!}{%
\begin{tabular}{l cc @{\hskip 10pt} ccc}
\toprule
& \multicolumn{2}{c}{\textbf{Cost (FPS $\uparrow$)}} & \multicolumn{3}{c}{\textbf{Responsiveness}} \\
\cmidrule(lr){2-3} \cmidrule(lr){4-6}
\textbf{Method} & \textbf{A6000} & \textbf{H100} & \textbf{Switched $\le$2 chunks $\uparrow$} & \textbf{Onset (s) $\downarrow$} & \textbf{$\Delta$CLIP $\uparrow$} \\
\midrule
Self Forcing   & 4.14 & 16.98 & 0.07 & 0.75 & 0.65 \\
\oursrow + \ours & 3.91 & 15.98 & \textbf{0.35} & \textbf{0.12} & \textbf{3.91} \\
\midrule
LongLive       & 3.52 & 14.02 & 0.21 & 0.62 & 1.63 \\
\oursrow + \ours & 3.31 & 13.11 & \textbf{0.54} & \textbf{0.19} & \textbf{4.22} \\
\bottomrule
\end{tabular}%
}
\vspace{-10pt}
\end{table}
Tab.~\ref{tab:practical} shows that RG makes switches take effect faster at a small cost.
RG adds one forward pass per chunk, which lowers end-to-end FPS by about 6\% for a 5\secs video with one switch on both an RTX A6000 and an H100.
The FPS in Tab.~\ref{tab:practical} is measured over the whole 5 s video, which includes VAE decoding and the three pre-switch chunks generated without RG.
In long sessions where RG is active on every chunk, the cost is bounded by the per-chunk overhead.
Measured on the post-switch chunks alone, the chunks generated with RG run at 21.7 and 18.2 FPS on Self Forcing and LongLive on the H100, which is 13\% and 16\% below the base models and still above the 16 FPS playback rate, and peak memory is unchanged.
We measure responsiveness by the fraction of switches realized within two chunks (\emph{Switched $\leq$2 chunks}), the increase in CLIP similarity to the new prompt relative to the old one after the switch ($\Delta$CLIP), and the time at which this increase reaches half of its final value (\emph{Onset}).
RG more than doubles both the switched fraction and $\Delta$CLIP on both models and brings the median onset below 0.2\secs.
Without RG, the onset stays at 0.62\secs even on LongLive, whose cache adaptation at a switch does not remove context inertia.

\subsection{Ablation Results}
\label{supp:ablation}

\begin{table*}[h!]
\centering
\caption{\textbf{Ablations.} 100 random pairs from the T2V-CompBench dynamic attribute set on Self Forcing (SF) and LongLive (LL). Default: $\tau = 850$, $\lambda = (1, 0.5, 0, 0)$, and one $\epsilon$ shared within the prompt interval. (a) Noise level $\tau$. (b) Step schedule $\lambda$. (c) Noise sharing.}
\label{tab:ablation}
\footnotesize
\setlength{\tabcolsep}{3pt}
\renewcommand{\arraystretch}{1.05}

\begin{minipage}[t]{0.31\textwidth}
\vspace{0pt}
\centering
\subcaption{Noise level $\tau$}
\label{tab:abl_tau}
\resizebox{\linewidth}{!}{%
\begin{tabular}{l cc cc cc}
\toprule
& \multicolumn{2}{c}{Achv.} & \multicolumn{2}{c}{Quality} & \multicolumn{2}{c}{Subject} \\
\cmidrule(lr){2-3} \cmidrule(lr){4-5} \cmidrule(lr){6-7}
$\tau$ & SF & LL & SF & LL & SF & LL \\
\midrule
none & 0.130 & 0.350 & 0.789 & 0.770 & 0.988 & 0.989 \\
700  & 0.260 & 0.425 & 0.791 & 0.778 & 0.975 & 0.984 \\
800  & 0.468 & 0.550 & 0.787 & 0.768 & 0.962 & 0.982 \\
\oursrow 850 & 0.623 & 0.675 & 0.788 & 0.778 & 0.957 & 0.979 \\
900  & 0.779 & 0.775 & 0.790 & 0.771 & 0.954 & 0.979 \\
1000 & 0.883 & 0.850 & 0.795 & 0.772 & 0.956 & 0.975 \\
\bottomrule
\end{tabular}%
}
\end{minipage}\hfill
\begin{minipage}[t]{0.34\textwidth}
\vspace{0pt}
\centering
\subcaption{Step schedule $\lambda$}
\label{tab:abl_steps}
\resizebox{\linewidth}{!}{%
\begin{tabular}{l cc cc cc}
\toprule
& \multicolumn{2}{c}{Achv.} & \multicolumn{2}{c}{Quality} & \multicolumn{2}{c}{Subject} \\
\cmidrule(lr){2-3} \cmidrule(lr){4-5} \cmidrule(lr){6-7}
$\lambda$ & SF & LL & SF & LL & SF & LL \\
\midrule
(1, 0, 0, 0)     & 0.468 & 0.625 & 0.789 & 0.774 & 0.963 & 0.981 \\
\oursrow (1, 0.5, 0, 0) & 0.623 & 0.675 & 0.788 & 0.778 & 0.957 & 0.979 \\
(1, 1, 0, 0)     & 0.753 & 0.838 & 0.778 & 0.771 & 0.942 & 0.969 \\
(1, 1, 1, 0)     & 0.805 & 0.850 & 0.753 & 0.738 & 0.866 & 0.896 \\
\bottomrule
\end{tabular}%
}
\end{minipage}\hfill
\begin{minipage}[t]{0.33\textwidth}
\vspace{0pt}
\centering
\subcaption{Noise sharing $\epsilon$}
\label{tab:abl_eps}
\resizebox{\linewidth}{!}{%
\begin{tabular}{l cc cc cc}
\toprule
& \multicolumn{2}{c}{Achv.} & \multicolumn{2}{c}{Quality} & \multicolumn{2}{c}{Subject} \\
\cmidrule(lr){2-3} \cmidrule(lr){4-5} \cmidrule(lr){6-7}
$\epsilon$ & SF & LL & SF & LL & SF & LL \\
\midrule
per chunk & 0.675 & 0.662 & 0.786 & 0.777 & 0.939 & 0.976 \\
\oursrow per switch & 0.623 & 0.675 & 0.788 & 0.778 & 0.957 & 0.979 \\
\bottomrule
\end{tabular}%
}
\end{minipage}

\end{table*}

\begin{figure}[t]
  \centering
  \includegraphics[width=\linewidth]{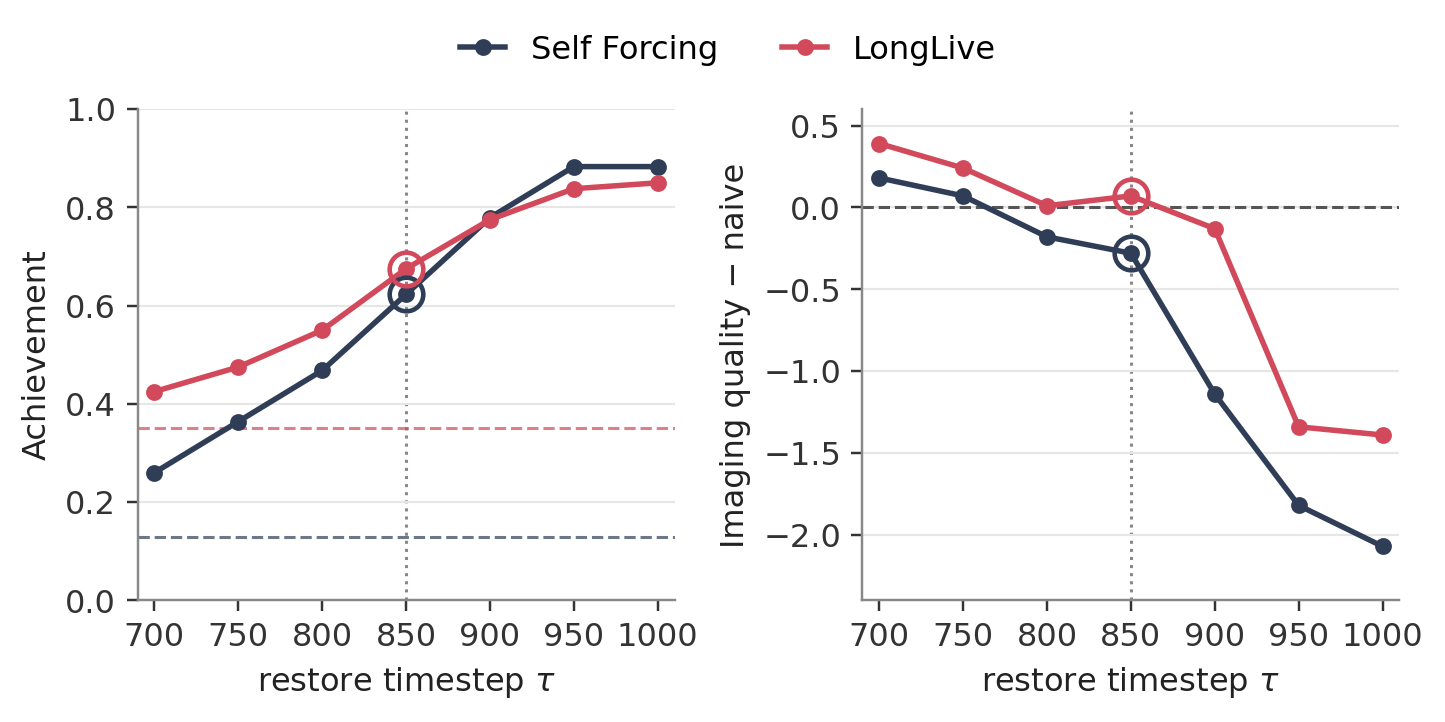}
  \caption{\textbf{On the choice of noise level $\tau$.} Achievement (left) and imaging quality relative to the base model (right) on 100 pairs. Dashed lines mark the base models. Achievement rises with $\tau$ throughout, whereas imaging quality holds up to $\tau = 850$ (circled) and drops sharply beyond, which is where the restored chunk no longer preserves the subject (Fig.~\ref{fig:ablation_qual}).}
  \label{fig:tau_sweep}
\end{figure}

We evaluate all ablations on 100 random pairs from the T2V-CompBench dynamic attributes set on Self Forcing (SF) and LongLive (LL). 
Unless otherwise specified, we use $\tau = 850$, $\lambda = (1, 0.5, 0, 0)$, and shared noise $\epsilon$, varying one factor at a time (Tab.~\ref{tab:ablation}).

\paragraph{Noise level.}
The noise level $\tau$ controls how far the restored chunk can depart from the previous one, and thus how strongly the new prompt overrides the context.
Achievement rises monotonically with $\tau$, from 0.26 at $\tau = 700$ to 0.88 at $\tau = 1000$ on Self Forcing, and LongLive follows the same trend (Tab.~\ref{tab:abl_tau}).
Imaging quality, in contrast, remains stable up to $\tau = 850$ but drops by more than one point beyond it (Fig.~\ref{fig:tau_sweep}).
Fig.~\ref{fig:ablation_qual} illustrates this trade-off: at $\tau = 700$, the sign remains yellow, at $\tau = 850$, its color changes while its shape is preserved, and at $\tau = 1000$, it is regenerated with a different shape, indicating that the restored chunk no longer preserves the subject.
We therefore use $\tau = 850$, where the subject survives the switch while allowing the requested change.

Importantly, $\tau$ is better viewed as a user-controllable knob than as a parameter that requires per-configuration tuning for RG to work.
Similar to the guidance scale in classifier-free guidance, it controls the trade-off between responsiveness to the new prompt and preservation of the current scene.
Indeed, every value in our sweep improves achievement over the base model.
For simplicity and consistency, we use the same $\tau = 850$ across all models and benchmarks, without per-configuration tuning.
Users can adjust it to favor either responsiveness or preservation, and the preferred balance may vary across models and prompt types.

\paragraph{Step schedule.}
Applying guidance only at the first step yields partial improvements in Achievement (0.47 and 0.63). 
Extending full-strength guidance to the second step further improves responsiveness, but slightly compromises subject consistency. 
Applying it through the third step provides only modest additional gains while substantially degrading both quality and subject consistency. 
Our default schedule, which applies half-strength guidance at the second step, offers a better balance: it reinforces the prompt switch while leaving the remaining steps to refine the generated content.

\paragraph{Noise sharing.}
The cache-free pass re-noises the previous chunk with a noise sample $\epsilon$, so the redrawn chunk, and hence the residual $r_k$, depends on $\epsilon$.
If a new $\epsilon$ is drawn for every chunk, each chunk in the prompt interval is pulled toward a different realization of the new prompt, which shows up as subject drift across chunks.
We therefore draw one $\epsilon$ at the switch and reuse it for every chunk in the prompt interval.
Tab.~\ref{tab:abl_eps} shows that sharing improves subject consistency, most clearly on Self Forcing, resulting in slightly higher Quality on both models.
Achievement is similar on LongLive and lower on Self Forcing, which we attribute to per-chunk resampling changing the subject more aggressively.

\section{Additional Experiments}
\label{supp:more_exp}

\begin{figure*}[h!]
\centering
\includegraphics[width=\textwidth]{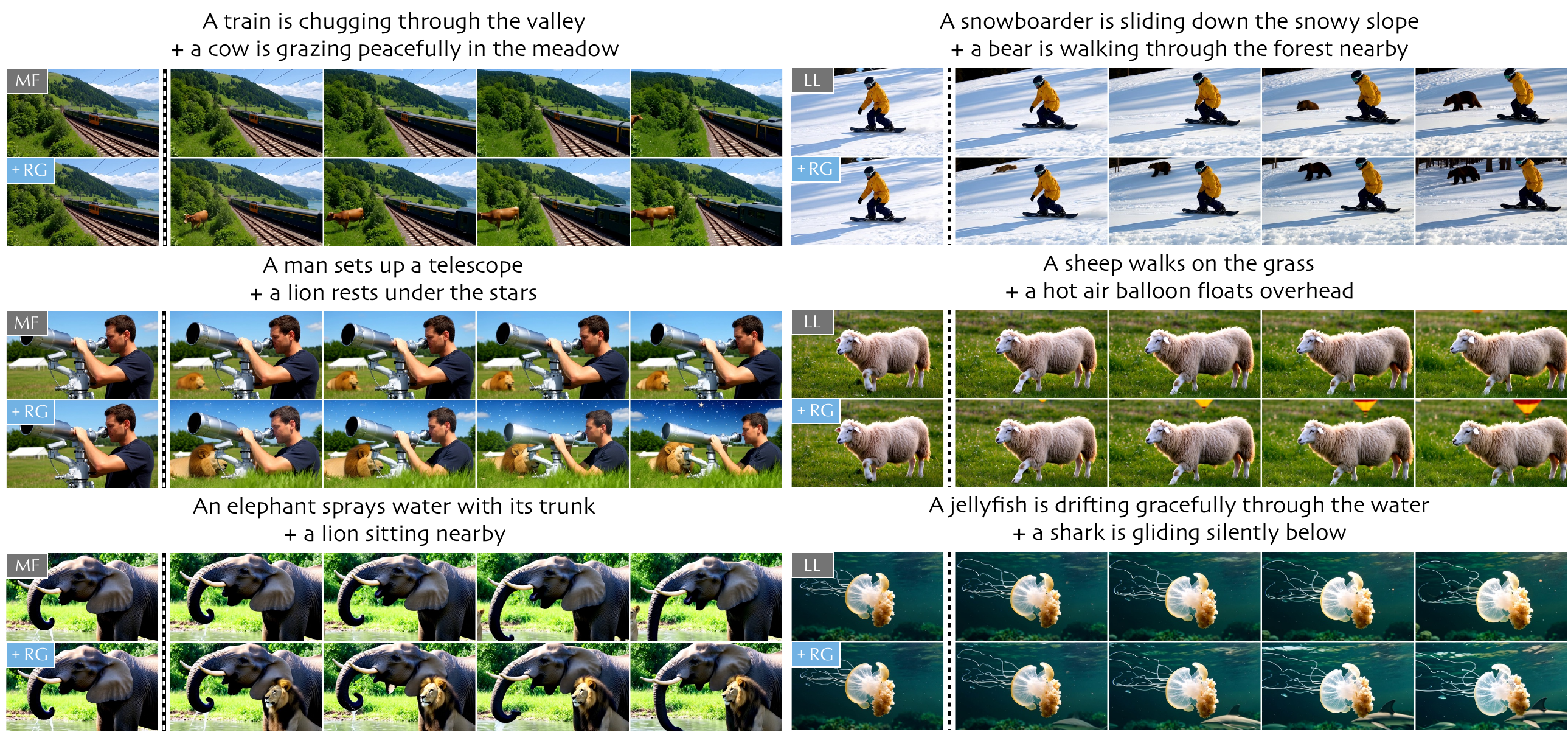}
\caption{\textbf{Qualitative results on T2V-CompBench action binding.} The white-to-black bar marks the prompt switch, at which a second subject is added to the prompt. Without RG the video keeps the initial scene, whereas with RG the new subject appears within the first chunks after the switch while the original subject and background are preserved.}
\label{fig:actionbind}
\end{figure*}
\begin{table*}[h!]
\centering
\caption{\textbf{Prompt switching on T2V-CompBench action binding.} These prompts involve adding a second subject at the switch. Metrics follow Tab.~\ref{tab:dynattr}, with the official action binding score of T2V-CompBench (normalized to $[0,1]$) in place of the dynamic attribute score.}
\label{tab:actionbind}
\small
\resizebox{0.75\textwidth}{!}{%
\begin{tabular}{l cccccc}
\toprule
\textbf{Method} & \textbf{Achievement $\uparrow$} & \textbf{T2V-CB Act.\ Bind.\ $\uparrow$} & \textbf{CLIP $\uparrow$} & \textbf{Dynamic $\uparrow$} & \textbf{Quality $\uparrow$} & \textbf{Overall $\uparrow$} \\
\midrule
MemFlow~\cite{ji2025memflow}
& 0.386 & 0.524 & 26.39 & 0.680 & 0.836 & 0.746 \\
\oursrow +\ours
& \underline{0.605} & \underline{0.624} & \underline{28.00} & \textbf{0.860} & \textbf{0.839} & \underline{0.793} \\
\midrule
LongLive~\cite{yang2025longlive}
& 0.450 & 0.564 & 27.24 & 0.665 & \underline{0.838} & 0.761 \\
\oursrow +\ours
& \textbf{0.667} & \textbf{0.627} & \textbf{28.56} & \underline{0.810} & 0.837 & \textbf{0.803} \\
\bottomrule
\end{tabular}%
}
\end{table*}

\subsection{T2V-CompBench Action Binding}
\label{supp:actbind}
The action binding prompts of T2V-CompBench describe two subjects performing different actions. We use the first subject as $\yold$ and append the second as $\ynew$, so that the switch asks the model to introduce a new subject into the scene rather than to change an attribute of the existing one. This is a different kind of switch from dynamic attribute: the previous content must be kept in full and a new element added beside it. We evaluate on MemFlow and LongLive, and report the official action binding score of the benchmark alongside the metrics of Tab.~\ref{tab:dynattr}.

Tab.~\ref{tab:actionbind} shows that RG raises Achievement from 0.386 to 0.605 on MemFlow and from 0.450 to 0.667 on LongLive, and the official action-binding score and CLIP rise with it. Quality stays within 0.003 of the base model and dynamic degree increases, so the new subject is added without freezing the video. Fig.~\ref{fig:actionbind} shows examples. The base models keep the initial scene after the switch, whereas with RG the new subject appears within the first chunks (e.g., the cow in the meadow, the bear beside the snowboarder, the lion next to the elephant) while the original subject, its action and the background are preserved. The hot air balloon in the sheep example appears small and late, which reflects the difficulty of adding an object that occupies little of the frame.

\subsection{A Special Case: Rolling Forcing}
\label{supp:rolling}
\begin{table*}[h!]
\centering
\caption{\textbf{Residual guidance on Rolling Forcing.} 100 random prompt pairs of T2V-CompBench dynamic attribute. Metrics follow Tab.~\ref{tab:dynattr}.}
\label{tab:rolling}
\small
\resizebox{0.79\textwidth}{!}{%
\begin{tabular}{l cccccc}
\toprule
\textbf{Method} & \textbf{Achievement $\uparrow$} & \textbf{T2V-CB Dyn.\ Attr.\ $\uparrow$} & \textbf{CLIP $\uparrow$} & \textbf{Dynamic $\uparrow$} & \textbf{Quality $\uparrow$} & \textbf{Overall $\uparrow$} \\
\midrule
Rolling Forcing~\cite{liu2026rolling}
& 0.214 & 0.042 & 25.58 & 0.925 & \textbf{0.827} & 0.704 \\
\oursrow +\ours
& \textbf{0.476} & \textbf{0.095} & \textbf{26.11} & \textbf{0.935} & 0.821 & \textbf{0.752} \\
\bottomrule
\end{tabular}%
}
\end{table*}
\begin{figure*}[h!]
\centering
\includegraphics[width=\textwidth]{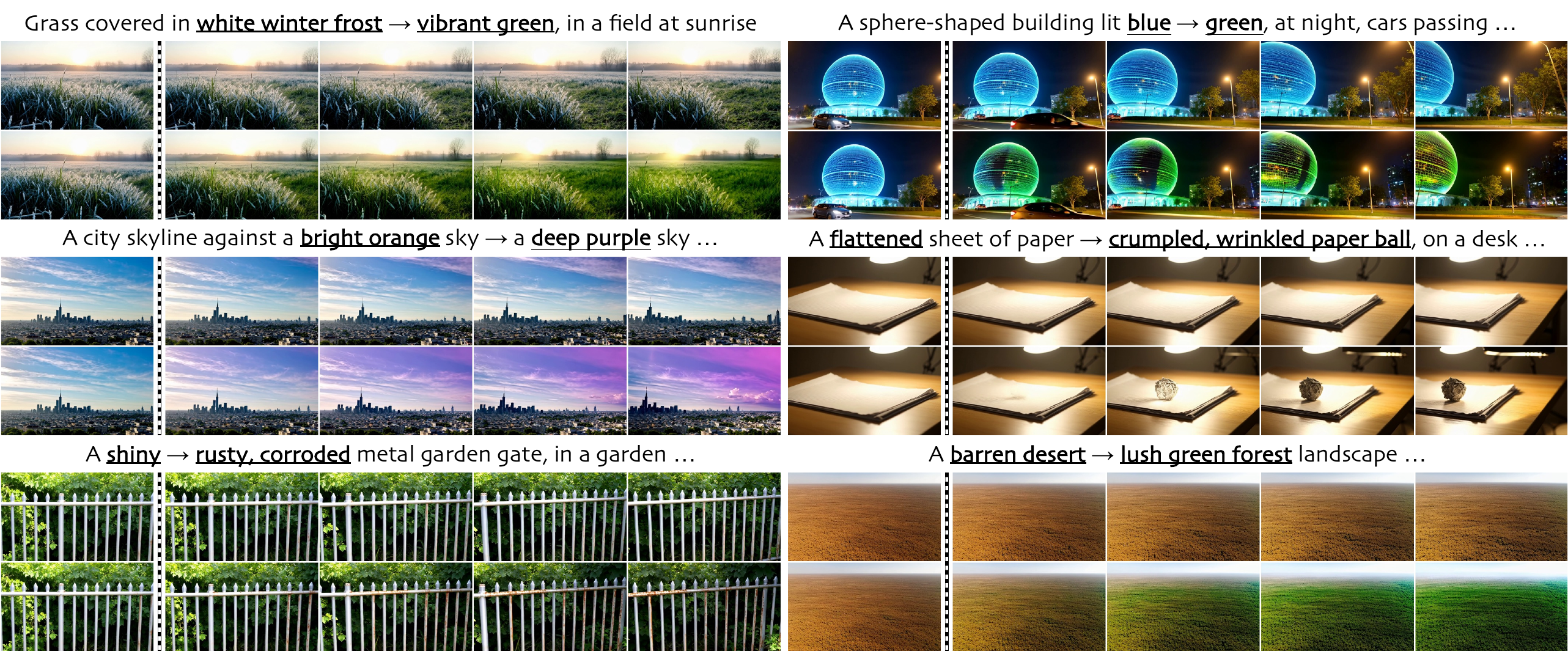}
\caption{\textbf{Qualitative results on Rolling Forcing.} T2V-CompBench dynamic attribute. In each example, the top row is Rolling Forcing and the bottom row is Rolling Forcing with RG. The white-to-black bar marks the prompt switch.}
\label{fig:rolling}
\end{figure*}

Unlike chunk-wise models, Rolling Forcing~\cite{liu2026rolling} denoises a window of $T{=}4$ chunks held at staggered noise levels and advances the whole window by one step per forward pass.
With each pass the front chunk leaves the window clean and is written to the KV cache, and a pure-noise chunk enters at the back.
A chunk therefore receives its first prediction when it enters the window, and the step index of Eq.~\eqref{eq:inject} corresponds to its position from the back.
Residual guidance is mapped onto this schedule as follows.
Before each post-switch chunk $k$ enters the window, $s_{k-1}$ is taken as the clean chunk that most recently left the window, which is also the last one written to the cache.
It is re-noised to $\tau{=}850$ and restored once under the new prompt without the cache, which gives $r_k$ as in Eq.~\eqref{eq:residual}.
The residual is then applied by window position: $\lambda_1 r_k$ is added to the prediction of chunk $k$ while it sits at the back of the window and $\lambda_2 r_k$ at the next position, so that every chunk receives the schedule of the chunk-wise setting exactly once as it passes through the window.
Chunks from the previous prompt that are still in the window at the switch are left unmodified, so the frames before the switch are preserved.
For the first post-switch chunk no chunk has yet left the window under the new prompt, and the prediction of the front chunk, which is the most refined one in the window, is used as $s_{k-1}$.
Tab.~\ref{tab:rolling} reports the result on T2V-CompBench dynamic attribute.
RG raises Achievement from 0.214 to 0.476 while Quality stays within 0.006 of the base model, and Overall increases from 0.704 to 0.752.
Fig.~\ref{fig:rolling} shows qualitative examples: the base model keeps the initial state after the switch, whereas RG realizes the change within the first chunks (e.g., the frost melts into green grass and the sheet of paper crumples into a ball) while the layout and camera remain continuous. 
Note that the switch is realized only after the chunks already in the window at the switch have left it, so the change appears one or two chunks later than in the chunk-wise models.

\subsection{Very Long Generation (240\secs)}
\label{supp:240s}

\begin{table}[t]
\centering
\caption{\textbf{240\secs generation on LongLive.} CLIP and VBench quality per minute for LongLive with and without residual guidance, averaged over 50 scripts.}
\label{tab:240s}
\footnotesize
\begin{tabular}{l cc cc}
\toprule
& \multicolumn{2}{c}{\textbf{CLIP $\uparrow$}} & \multicolumn{2}{c}{\textbf{Quality $\uparrow$}} \\
\cmidrule(lr){2-3} \cmidrule(lr){4-5}
\textbf{Minute} & LongLive & +\ours & LongLive & +\ours \\
\midrule
0--1 & 27.34 & \textbf{27.90} & \textbf{0.813} & 0.811 \\
1--2 & 26.67 & \textbf{27.43} & \textbf{0.808} & 0.804 \\
2--3 & 25.95 & \textbf{27.15} & \textbf{0.810} & 0.805 \\
3--4 & 25.84 & \textbf{26.99} & \textbf{0.809} & 0.801 \\
\bottomrule
\end{tabular}
\end{table}

To test whether the gain persists over long rollouts, we extend 50 InteractiveBench scripts from 60\secs to 240\secs. Each script keeps its six original prompts and is continued with 18 further prompts written by Qwen3-VL-32B, which keep the setting, characters and visual style, introduce one visible event or state change per prompt, and re-describe the scene in full as in the original benchmark. A video therefore spans 24 prompts of about 10\secs and 23 switches. We generate with LongLive, with and without residual guidance, under the configuration of Sec.~\ref{sec:exp}, and both conditions share the seed of the 60\secs run. Following LongLive~\cite{yang2025longlive}, we report CLIP per prompt segment, grouped by minute, together with the VBench quality score.

Tab.~\ref{tab:240s} shows that the gain in CLIP holds throughout the video and grows over time, from $+0.56$ in the first minute to $+1.15$ in the last. The base model drifts further from each new prompt as the video lengthens, falling from 27.34 to 25.84, whereas residual guidance keeps following the prompt and falls only to 26.99. Quality stays within 0.008 of the base model in every minute, indicating that repeated prompt switches introduce little additional degradation in the measured quality dimensions.

\section{Discussion}
\label{supp:discussion}

\subsection{Relation to Prior Work}
\label{supp:rwork}
\begin{table*}[h!]
\centering
\caption{\textbf{Comparison with existing guidance-based methods.} Classifier-free guidance (CFG), History guidance (HG) and the guidance of MAGI-1 contrast predictions of the same noisy sample at every step. Our Residual guidance (RG) contrasts a context-free prediction against the previous clean chunk, once per chunk, and needs no unconditional branch or training.}
\label{tab:guidance_compare}
\resizebox{0.75\textwidth}{!}{%
\begin{tabular}{l|cccc}
\hline
 & CFG~\cite{ho2022classifier} & HG~\cite{song2025history} & MAGI-1~\cite{teng2025magi} & RG (ours) \\
\hline
Strengthens & prompt & context & context and prompt & prompt over context \\
Contrast    & prompt vs.\ none & context vs.\ none & three branches & no context vs.\ $s_{k-1}$ \\
Reference   & noisy sample & noisy sample & noisy sample & clean chunk \\
Injection   & every step & every step & every step & early steps \\
Control     & scale $w$ & scale $w$ & scales $w_{\text{prev}}, w_{\text{text}}$ & noise level $\tau$ \\
Requires    & uncond.\ branch & dropout training & uncond.\ branch & none \\
\hline
\end{tabular}
}
\end{table*}
\begin{figure*}[h!]
  \centering
  \begin{subfigure}[b]{0.49\textwidth}
    \centering
    \includegraphics[width=\linewidth]{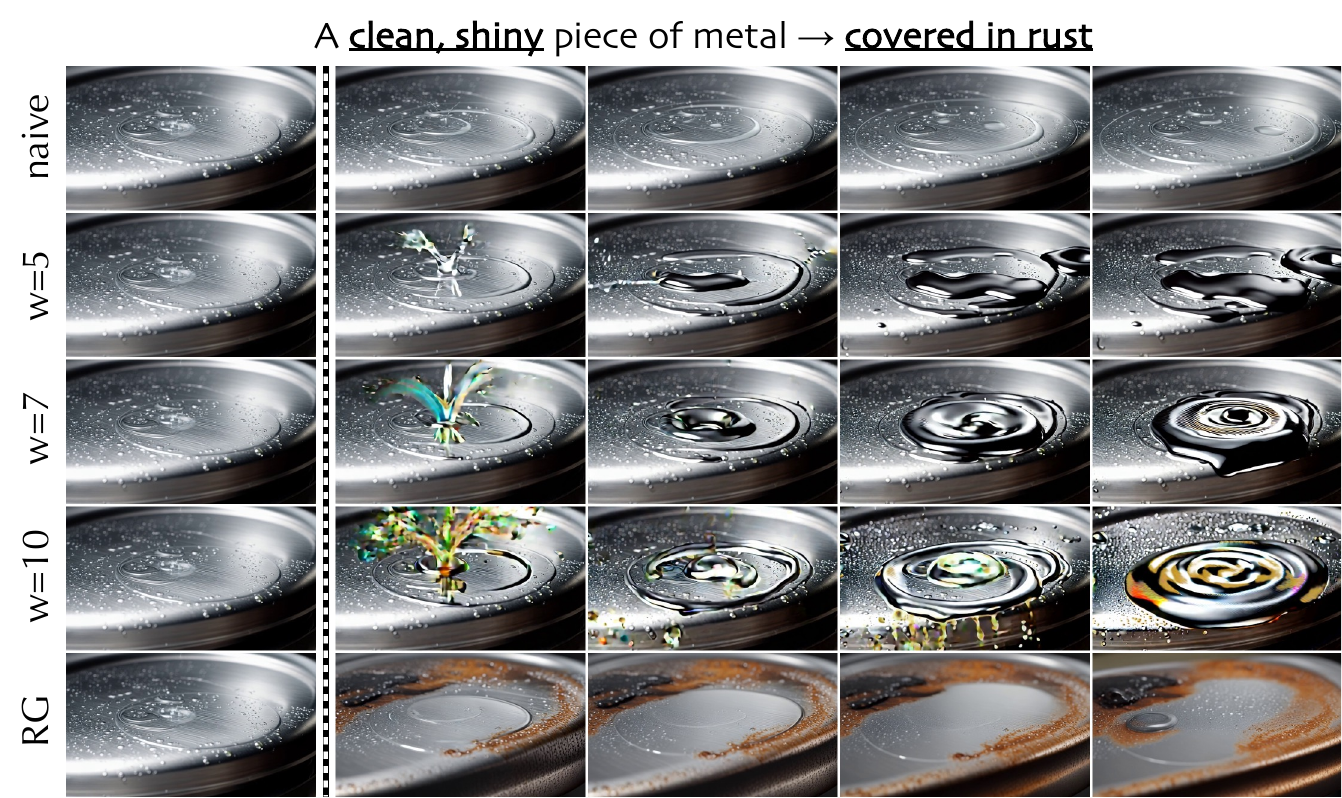}
    \caption{Self Forcing}
    \label{fig:cfg_sf}
  \end{subfigure}\hfill
  \begin{subfigure}[b]{0.49\textwidth}
    \centering
    \includegraphics[width=\linewidth]{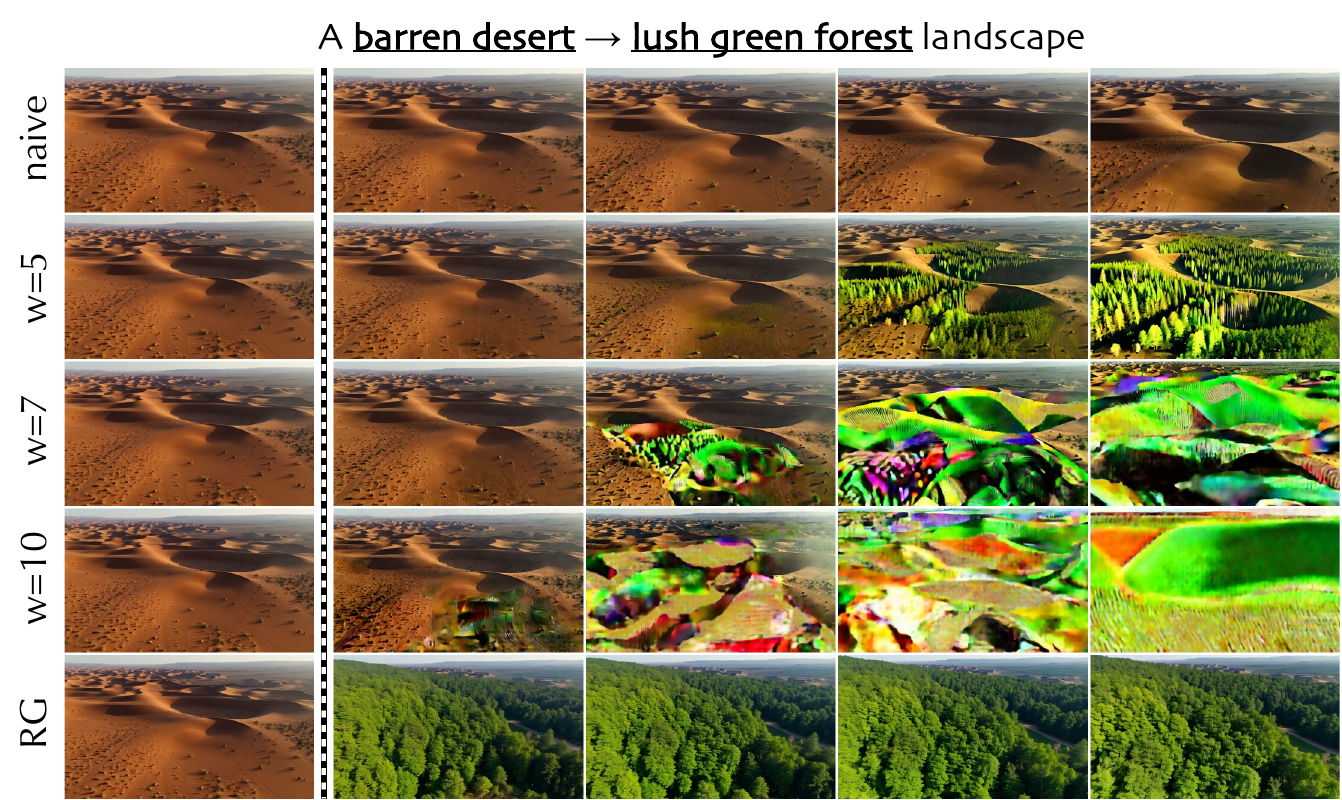}
    \caption{LongLive}
    \label{fig:cfg_ll}
  \end{subfigure}
  \caption{\textbf{Classifier-free guidance on the switch chunk.}
     Applying CFG on (a) Self Forcing and (b) LongLive. \emph{Naive} denotes the base model. For both models, CFG partially enables the switch but causes artifacts while Residual guidance realizes the switch without artifacts. }
  \label{fig:cfg}
\end{figure*}
\paragraph{Relation to prior analyses.}
Several recent works report related forms of history-induced suppression, but approach them primarily through cache or memory design.
LongLive~\cite{yang2025longlive} observes that retaining the KV cache delays or prevents prompt switches, attributes this to residual information from the previous prompt, and introduces KV recaching to address it.
SWIFT~\cite{tan2026swift} similarly identifies a mismatch between cached video history and updated prompt semantics, and proposes semantic injection and adaptive memory allocation.
Echo-Forcing~\cite{wu2026echo} identifies interference from outdated scene memories and addresses it through structured scene memories and discrepancy-aware memory decay.
In the related but distinct setting of action-conditioned world modeling, Astra~\cite{zhu2026astra} studies how longer visual histories reduce action responsiveness, refers to this phenomenon as \emph{visual inertia}, and mitigates it through noise-augmented history memory.

A second line of work examines the denoising steps of autoregressive video diffusion under a fixed prompt.
Bai et al.~\cite{bai2026causality} find that content and motion are largely established at the first denoising step and attribute this to the historical context.
HiAR~\cite{zou2026hiar} relate the first step of few-step students to motion dynamics and sample diversity.
In-Context Forcing~\cite{yang2026context} further shows that Self Forcing attends more strongly to the previous frame than to the current one at the first step, and links this to weakened dynamics.
These studies focus on motion, diversity, or temporal drift rather than prompt-switch responsiveness, and primarily address their findings through student retraining.

Our analysis complements both lines.
We isolate how the visual content of the retained context suppresses prompt-induced prediction changes, even after recaching under the new prompt.
We further measure the prompt response at each denoising step and show that the step at which the prompt acts most strongly is also the step at which the context suppresses it most, so that later steps cannot recover the switch.
The suppression is already substantial with a single preceding chunk.
Together, these findings identify where a training-free correction should act, which motivates the design of RG.

\paragraph{Relation to guidance-based methods.}

Tab.~\ref{tab:guidance_compare} compares residual guidance with the three guidance schemes that it most resembles, and all four can be written as a correction added to the prediction of the switch chunk. The key distinction lies in the reference used to construct each correction: CFG and MAGI-1 measure prompt directions under cached context, History Guidance reinforces the context, whereas RG estimates the new-prompt displacement without cached context.

Specifically, Classifier-free guidance~\cite{ho2022classifier} adds $(w-1)[\hat{x}(\mathcal{C}, y) - \hat{x}(\mathcal{C}, \varnothing)]$, the prompt direction measured under the context, at every step.
History guidance~\cite{song2025history} adds $(w-1)[\hat{x}(\mathcal{C}, y) - \hat{x}(\varnothing, y)]$, the context direction, which strengthens the context rather than the prompt, and it requires a model trained with history dropout so that $\hat{x}(\varnothing, y)$ is defined.
MAGI-1~\cite{teng2025magi} combines both with separate scales, $(w_{\text{text}}-1)[\hat{x}(\mathcal{C}, y) - \hat{x}(\mathcal{C}, \varnothing)] + (w_{\text{prev}}-1)[\hat{x}(\mathcal{C}, \varnothing) - \hat{x}(\varnothing, \varnothing)]$, and raises $w_{\text{prev}}$ to reduce flicker between chunks, which strengthens exactly the context that suppresses a switch.
Its prompt direction is the same as in classifier-free guidance, measured under the context, and it is disabled in the distilled MAGI-1 since the student has no null prompt.
Sec.~\ref{subsec:cause} shows that this direction shrinks under the context, so that the scale required to realize a switch saturates the prediction before the switch is reached. We discuss this more in Supp.~\ref{supp:cfg}.

Residual guidance differs in the quantity it adds. 
The direction $r_k = \hat{x}(\varnothing, \ynew) - s_{k-1}$ that we use is measured without the context and is therefore not a rescaling of the prompt direction of CFG and MAGI-1, which is measured under the context.
History guidance also evaluates a context-free prediction, but on the current noisy sample and as the point to move away from, which reinforces the context, whereas RG evaluates it on the re-noised previous chunk and moves the prediction toward it.
Its reference is the previous clean chunk rather than the current noisy sample, so it is computed once per chunk instead of at every step. Its magnitude is governed by the noise level $\tau$, while its contribution at each denoising step is controlled by $\lambda_i$, rather than by a CFG scale.
The context-free query is obtained from the student itself, which runs without a cache on the first chunk of every video, so no unconditional branch and no training are needed.

\subsection{CFG on Distilled Students}
\label{supp:cfg}
With the context, the magnitude of the classifier-free direction
$\|\hat{x}(\mathcal{C}, \ynew) - \hat{x}(\mathcal{C}, y_\varnothing)\|$ at the first step of the switch chunk
is 3 to 6 times smaller than its context-free counterpart
(4.7$\times$ on average over five models, Tab.~\ref{tab:cfg_shrink}).
Amplifying this direction therefore requires a large scale before the prompt can take effect.

\begin{table}[t]
\centering
\caption{\textbf{The context shrinks the classifier-free direction.}
$\|\hat{x}(\mathcal{C}, \ynew) - \hat{x}(\mathcal{C}, y_\varnothing)\| / \|s_{k-1}\|$
at the first denoising step of the switch chunk, without context ($\mathcal{C}=\varnothing$)
and with the full context, on the 100 switch pairs of Sec.~\ref{sec:analysis}
(mean, 95\% bootstrap CI). LongLive is measured after re-caching.}
\label{tab:cfg_shrink}
\resizebox{\columnwidth}{!}{%
\begin{tabular}{lccc}
\toprule
Model & w/o context & w/ context & Ratio \\
\midrule
Self Forcing    & 0.907 {\scriptsize [0.882, 0.933]} & 0.157 {\scriptsize [0.140, 0.175]} & 5.8$\times$ \\
Causal Forcing  & 0.774 {\scriptsize [0.746, 0.801]} & 0.253 {\scriptsize [0.228, 0.277]} & 3.1$\times$ \\
LongLive        & 0.892 {\scriptsize [0.869, 0.915]} & 0.182 {\scriptsize [0.164, 0.200]} & 4.9$\times$ \\
MemFlow         & 0.926 {\scriptsize [0.901, 0.950]} & 0.190 {\scriptsize [0.171, 0.210]} & 4.9$\times$ \\
Infinity-RoPE   & 0.827 {\scriptsize [0.806, 0.848]} & 0.176 {\scriptsize [0.155, 0.195]} & 4.7$\times$ \\
\bottomrule
\end{tabular}%
}
\end{table}

A natural alternative to residual guidance is to amplify the prompt direction that the student already provides, $\hat{x}(\mathcal{C}, \ynew) - \hat{x}(\mathcal{C}, y_\varnothing)$, with a scale $w$ as in classifier-free guidance.
This direction is measured under the context.
At the first step of the switch chunk, its magnitude is 3 to 6 times smaller than that of its context-free counterpart (4.7$\times$ on average over five models, Tab.~\ref{tab:cfg_shrink}), so the scale required to move the prediction as far as a switch demands is large.
Fig.~\ref{fig:cfg} shows the result for $w \in \{5, 7, 10\}$ on Self Forcing and LongLive.
At $w{=}5$ the switch is only partially realized, as the metal remains mostly clean and the forest appears only at the edges of the desert.
At $w{=}7$ and $w{=}10$ the prediction is pushed far enough to change, but the same scale amplifies every component of the direction, including the ones unrelated to the switch, and the frames show saturated colors and broken geometry before the new prompt is reached.
No scale in between realizes the switch cleanly, because the useful part of the direction and the artifacts grow together.
Residual guidance instead adds a direction that is measured without the context and is therefore large enough without amplification, so the switch is realized without the artifacts of a large guidance scale while the scene is largely retained.

\begin{figure}[t]
  \centering
  \includegraphics[width=\linewidth]{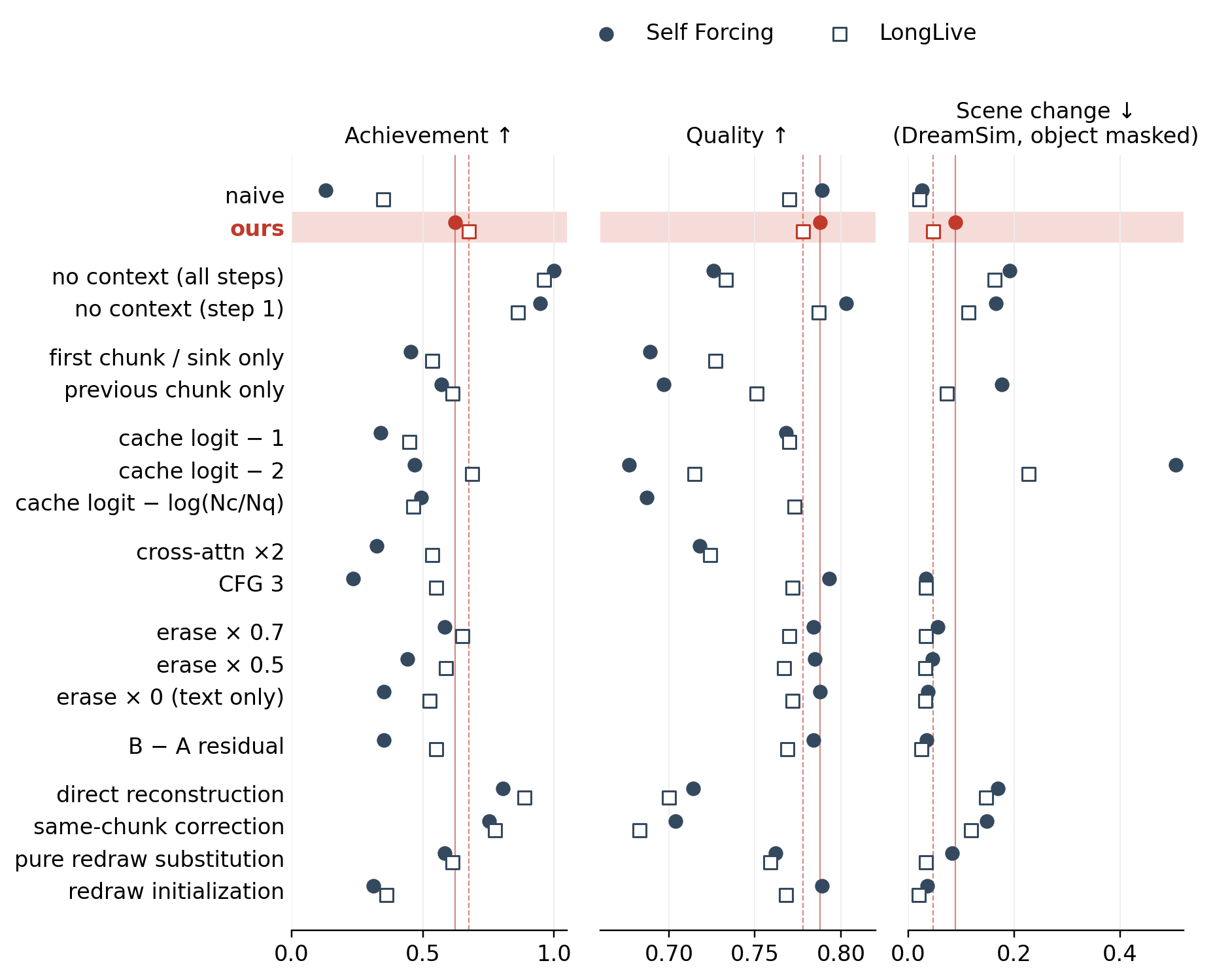}
  \caption{\textbf{Alternatives to residual guidance.} Achievement, Quality and scene change (object-masked DreamSim) across the switch for the base model, residual guidance (shaded) and alternatives that weaken the context, strengthen the prompt, change the erase strength, or use the cache-free reconstruction in other ways (prompt-difference (B$-$A) residual, direct reconstruction, same-chunk correction, pure substitution, redraw initialization), on 100 dynamic attribute pairs with Self Forcing (circles) and LongLive (squares). Vertical lines mark residual guidance. Scene change was not measured for the variants without a marker in the right panel. \emph{Naive} denotes the base model.}
  \label{fig:alternatives}
\end{figure}

\begin{figure*}[h!]
  \centering
  \includegraphics[width=\textwidth]{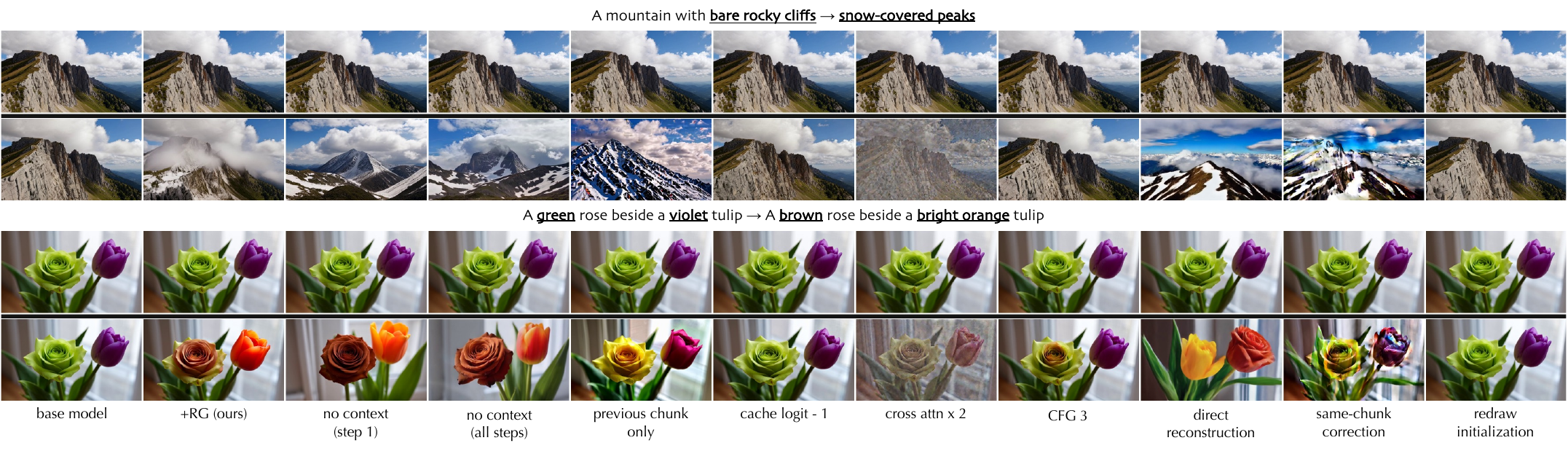}
  \caption{\textbf{Qualitative comparison of the alternatives.} For each switch, the top row shows a frame before the switch and the bottom row a frame from the last chunk after it, on Self Forcing (top) and LongLive (bottom). Only RG shows the new prompt while keeping the scene. Removing the context or reconstructing the chunk directly redraws the whole scene, previous chunk only changes the scene or the wrong attribute, cross-attention scaling and same-chunk correction add artifacts, and cache logit, CFG and redraw initialization keep the old state. See Fig.~\ref{fig:alternatives} for the quantitative comparison.}
  \label{fig:alternatives_quali}
\end{figure*}

\begin{figure*}[h!]
  \centering
  \includegraphics[width=\linewidth]{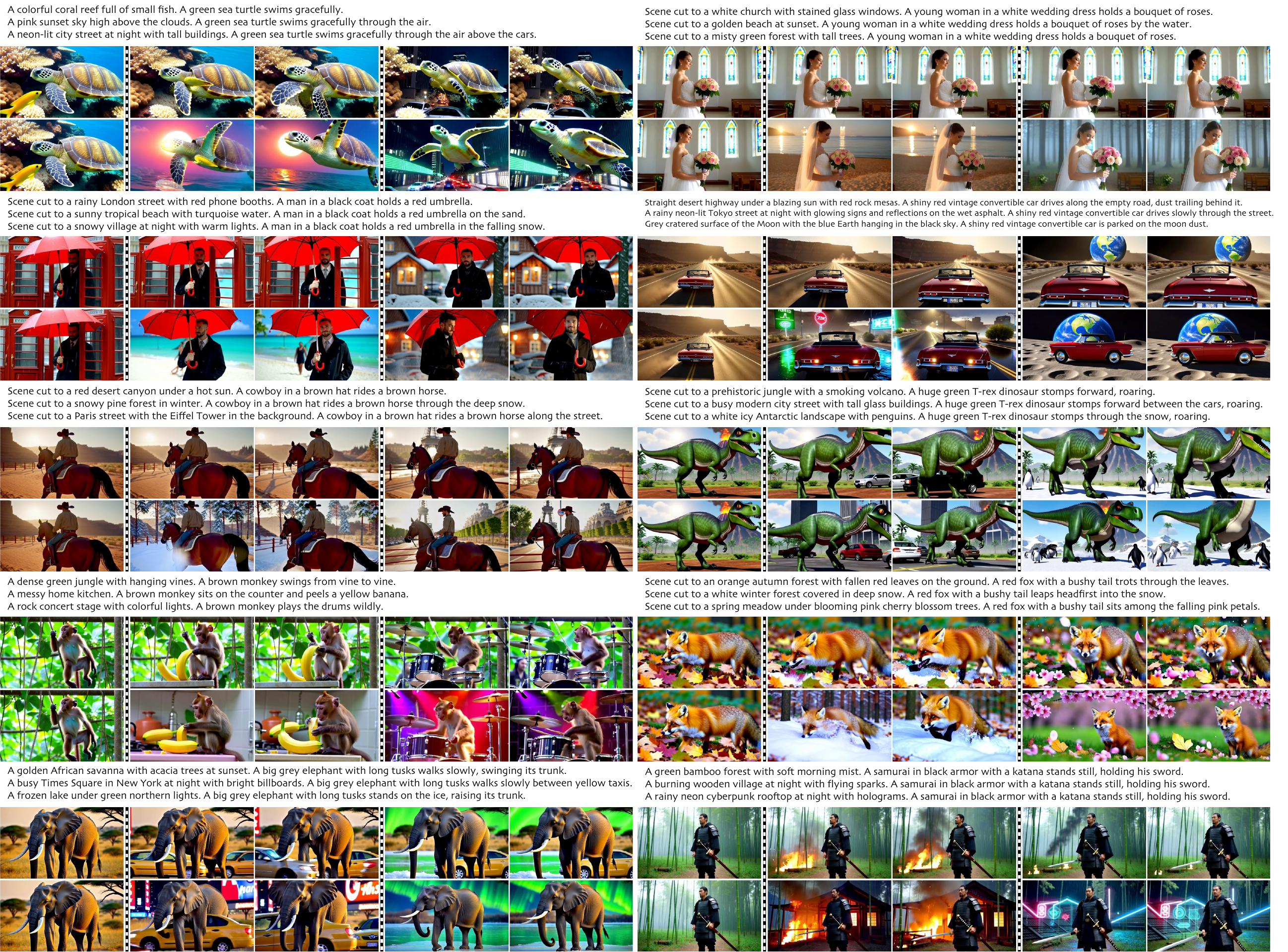}
  \caption{\textbf{Scene cuts on Infinity-RoPE.}
  Each panel shows three consecutive prompts that keep the subject and replace the scene.
  The top row is Infinity-RoPE and the bottom row is Infinity-RoPE with RG, generated from the same seed.
  Dashed lines mark the prompt switches.
  Infinity-RoPE often remains in the first scene or partially updates the scene,
  whereas RG moves to each new scene at the switch while retaining the subject.}
  \label{fig:scene_cut}
\end{figure*}

\subsection{On the Alternatives}
\label{supp:alternatives}

\paragraph{Setup.}
Sec.~\ref{sec:analysis} raises the question of whether the context can be weakened in other ways than adding a residual.
We compare alternatives on 100 randomly selected pairs of the T2V-CompBench dynamic attribute set with Self Forcing and LongLive (Fig.~\ref{fig:alternatives}).
All alternatives act on the same chunks as residual guidance (RG), from the switch chunk to the end of the video, at every denoising step unless noted.
Besides Achievement and Quality, we report the scene change across the switch: the DreamSim distance between the frames before and after the switch with the changed object masked out, so a method is penalized for altering the background and not for realizing the prompt.

\paragraph{Weakening the context.}
\emph{No context} removes self-attention to the cache, either at every step or at the first step only.
\emph{First chunk only} and \emph{previous chunk only} cut the cache down to the first chunk (the sink for LongLive) or to the chunk right before the current one.
\emph{Attention down-weighting} subtracts a constant $b$ from the attention logits of the cached keys, with $b \in \{1, 2\}$ and $b = \log(N_{\text{cache}} / N_{\text{cur}})$, which balances the attention mass of the cache against the current chunk.

\paragraph{Strengthening the prompt.}
\emph{Cross-attention $\times 2$} doubles the output of the text cross-attention.
\emph{CFG 3} applies classifier-free guidance to the distilled student, $\hat{x}(\mathcal{C}, \ynew) + 2\,[\hat{x}(\mathcal{C}, \ynew) - \hat{x}(\mathcal{C}, y_\varnothing)]$.

\paragraph{Variants of residual guidance.}
Let $\hat{s}(y) = \hat{x}_\theta(\tilde{x}_{k-1}^{\tau} \mid \varnothing, y)$ denote the cache-free reconstruction of the preceding chunk under prompt $y$, so that RG adds $r_k = \hat{s}(\ynew) - s_{k-1}$ (Eq.~\eqref{eq:residual}).
\emph{Erase scaling} splits this residual into a text term and an erase term,
$r_k = [\hat{s}(\ynew) - \hat{s}(y_\varnothing)] + \gamma\,[\hat{s}(y_\varnothing) - s_{k-1}]$, with $\gamma \in \{0, 0.5, 0.7\}$.
Setting $\gamma = 1$ recovers RG, and $\gamma = 0$ keeps only the text term.
The \emph{prompt-difference residual} (B$-$A) replaces $s_{k-1}$ with the reconstruction under the previous prompt, $r_k = \hat{s}(\ynew) - \hat{s}(\yold)$.
The last four variants use the cache-free reconstruction $\hat{s}(\ynew)$ other than as a residual.
\emph{Direct reconstruction} outputs $\hat{s}(\ynew)$ itself as the guided chunk.
\emph{Same-chunk correction} adds $r_k$ to a reconstruction made with the context visible instead of to the current chunk.
\emph{Pure substitution} moves the prediction toward the reconstruction, $\hat{x} + \lambda_i\,[\hat{s}(\ynew) - \hat{x}]$ with $\hat{x} = \hat{x}_\theta(x_k^{t_i} \mid \mathcal{C}_{<k}, \ynew)$, using the step schedule $\lambda$ of RG.
\emph{Redraw initialization} starts the chunk from $\hat{s}(\ynew)$ re-noised to $\tau$, with the cache kept and no residual.

\paragraph{Results.}
Removing the context raises Achievement above that of RG, but at the cost of scene preservation.
When the context is removed at every step, Quality drops and the scene change becomes about two to 3.5 times as large as with RG.
When it is removed only at the first step, Quality is maintained, but the scene change is still about twice that of RG.
Keeping a single chunk, whether the first or the previous one, falls short of RG in both Achievement and Quality, and the previous chunk alone also changes the scene more than RG.
Attention down-weighting is unstable. With $b = 1$ Achievement barely moves, with $b = 2$ it matches RG on LongLive but breaks the scene on Self Forcing, and the balanced value sits in between and loses Quality.
Doubling the cross-attention and applying CFG each raise Achievement by less than 0.2, and doubling the cross-attention also costs Quality.

Scaling down the erase term lowers Achievement monotonically.
With the text term alone ($\gamma = 0$), Achievement reaches only 0.351 on Self Forcing and 0.525 on LongLive, which recovers about half of the gain of RG over the base model (Tab.~\ref{tab:alternatives}).
The prompt-difference residual (B$-$A), which also keeps only a prompt direction, likewise stays well below RG.
A prompt direction alone is therefore insufficient even when measured without the cache, and erasing what the context imposes contributes as much as strengthening the prompt.
A smaller $\gamma$ reduces the scene change at equal Quality, and $\gamma = 0.7$ comes close to RG in Achievement, but any $\gamma \neq 1$ introduces an additional hyperparameter and an extra empty-prompt pass.
At $\gamma = 1$, the empty-prompt terms cancel, so RG requires only one extra pass and no additional weight.

How the reconstruction enters the chunk matters as much as the reconstruction itself.
Outputting the cache-free reconstruction directly or correcting a context-visible one raises Achievement but loses as much Quality as removing the context, and the scene change rises accordingly.
Starting from the reconstruction keeps the scene but barely changes the content, as the cache pulls the chunk back to the previous state.
Pure substitution falls below RG in both Achievement and Quality, while adding the same information as a residual keeps the detail of the context-aware prediction.
RG is thus the simplest variant that raises Achievement substantially while keeping Quality at the level of the base model and the scene change small.

\paragraph{Qualitative comparison.}
Fig.~\ref{fig:alternatives_quali} shows two switches: on Self Forcing, a bare rocky cliff should become snow-covered, and on LongLive, a green rose and a violet tulip should turn brown and bright orange.
RG adds snow to the same slope, with clouds now covering the ridge, and recolors the same two flowers, while the camera and the layout stay as before the switch.
Without context, whether removed at the first step or at every step, the new prompt is reached, but a different mountain and a different bouquet are drawn.
Direct reconstruction also redraws the scene, and on the flowers it swaps their positions.
The previous chunk alone replaces the mountain with another snowy peak and colors the flowers yellow and magenta.
Doubling the cross-attention and same-chunk correction leave visible artifacts.
Lowering the cache logits by 1 and redraw initialization keep both videos in the old state, and CFG at $w{=}3$ changes only the center of the rose.
Among these variants, RG is the only one that follows the new prompt and largely keeps the scene, which matches the trade-off in Fig.~\ref{fig:alternatives}.

\subsection{Scene cuts}
\label{supp:scene_cuts}

Some prompts in our benchmarks already involve a scene cut, as in film editing.
For example, in Fig.~\ref{fig:supp_storyeval}, spilled water forms a river and a boat then sails on it.
Here, we examine this setting directly with prompts that keep the subject and replace the entire scene.
RG is applied to Infinity-RoPE with RoPE Cut enabled, which is its mechanism designed for scene cuts.
As shown in Fig.~\ref{fig:scene_cut}, Infinity-RoPE often remains in the previous scene even with RoPE Cut, whereas RG moves to the new scene at each switch while retaining the subject.
Since a scene cut requires a larger change than an in-scene switch, we use a higher noise level of $\tau=950$, which we find to replace the scene more reliably than the default $\tau=850$.

\subsection{Limitations and Future Work}
\label{supp:limitations}

\begin{figure*}[h!]
  \centering
  \includegraphics[width=\textwidth]{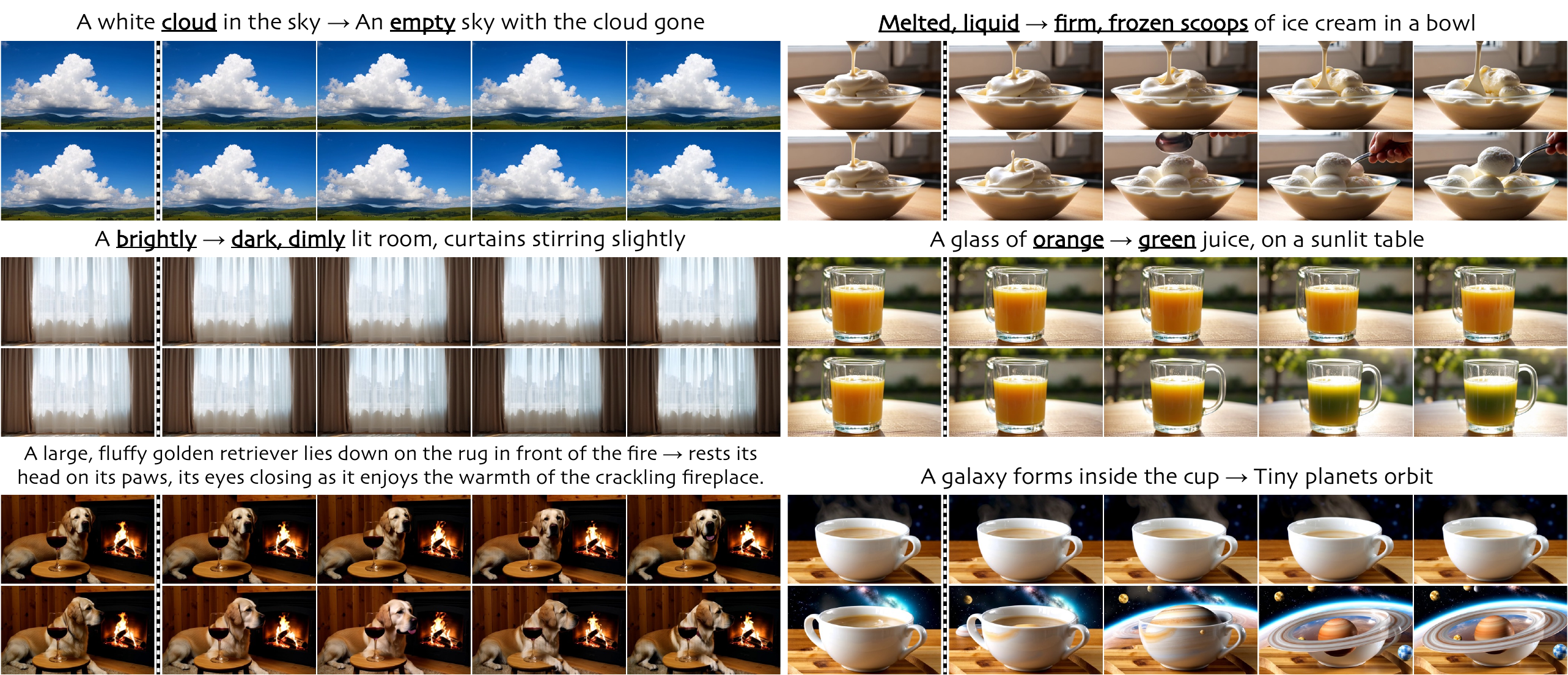}
  \caption{\textbf{Failure cases.}
  In each example, the top row is the base model and the bottom row is \ours. The first frame is the last frame before the switch, and the dashed bar marks the prompt switch.
  \textbf{Left: insufficient prompt response.} The requested removal of an object (LongLive), change of illumination (SWIFT) or change of pose (MemFlow, InteractiveBench) is not realized.
  \textbf{Right: unintended changes.} The switch is realized, but a spoon and a hand appear (Anchor Forcing), the glass acquires a handle (Self Forcing), or the cup itself turns into a planet (Anchor Forcing, StoryEval).}
  \label{fig:limitation}
\end{figure*}

Fig.~\ref{fig:limitation} shows representative failures of \ours, which fall into two types.

\paragraph{Insufficient prompt response.}
In the left column, the switch is not realized with or without \ours.
The cloud remains in the sky, the room becomes only slightly darker, and the dog does not rest its head on its paws.
These prompts request the removal of an object, a change of global illumination, or a change of fine-grained pose, while the layout of the scene stays the same.
We attribute these failures to the way the residual is obtained.
The previous chunk is re-noised only to an intermediate level, so its coarse structure and overall color survive the reconstruction, and the resulting residual is small for changes of this kind.
In particular, an object that is removed with nothing taking its place belongs to this coarse structure and tends to remain.
A higher noise level strengthens the response but also alters the scene (Sec.~\ref{sec:ablation}).
The third example comes from InteractiveBench, where consecutive prompts share most of their description, which further reduces the difference that the new prompt can induce.

\paragraph{Unintended changes.}
In the right column, the switch is realized but is accompanied by changes that the prompt does not request.
The ice cream becomes firm scoops, but a spoon and a hand enter the frame.
The juice turns green, but the glass acquires a handle.
The planets appear, but the cup itself turns into a ringed planet.
Since the residual is estimated without the cache, content that the new prompt leaves unspecified is not constrained by the preceding frames in this estimate, and the prior of the model can enter the next chunk through the residual.
This is most visible in the third example, where the new prompt does not mention the cup.

Together, these examples suggest a remaining trade-off between responsiveness and preservation: a weak correction can miss subtle changes, while an insufficiently selective correction can alter content beyond the prompt.
Future work could explore prompt-aware adaptation of the residual's strength and spatial extent to better balance these two objectives.

\end{document}